\documentclass[12pt]{article}

\usepackage{epsfig}
\usepackage{amssymb}

\newcommand{\beq}{\begin{equation}}
\newcommand{\bea}{\begin{eqnarray}}
\newcommand{\eeq}{\end{equation}}
\newcommand{\eea}{\end{eqnarray}}

\newcommand\Tr{\rm{Tr}\,}

\newcommand{\e}{{\rm e\,}}

\begin{document}

\title{\bf Real-time gauge theory simulations from
stochastic quantization with optimized updating}

\author{
J{\"u}rgen Berges\thanks{email: juergen.berges@physik.tu-darmstadt.de}
 \ and D{\'e}nes Sexty\thanks{email: denes.sexty@physik.tu-darmstadt.de}
\\[0.5cm]
Institute for Nuclear Physics\\
Darmstadt University of Technology\\
Schlossgartenstr. 9, 64289 Darmstadt, Germany}

\date{}
\begin{titlepage}
\maketitle
\def\thepage{}          

\begin{abstract}
We investigate simulations for gauge theories on a Minkowskian
space-time lattice. We employ stochastic quantization with
optimized updating using stochastic reweighting or gauge fixing,
respectively. These procedures do not affect the underlying theory
but strongly improve the stability properties of the stochastic
dynamics, such that simulations on larger real-time lattices can be
performed.
\end{abstract}

\end{titlepage}

\renewcommand{\thepage}{\arabic{page}}

\section{Introduction}

First-principles simulations for gauge field theories such as
quantum chromodynamics (QCD) on a Minkowskian space-time lattice
represent one of the outstanding aims of current research.
Typically, calculations are based on a Euclidean formulation,
where the time variable is analytically continued to imaginary
values. By this the quantum theory is mapped onto a statistical
mechanics problem, which can be simulated by importance sampling
techniques. In contrast, for real times standard importance
sampling is not possible because of a non-positive definite
probability measure.

Simulations in Minkowskian space-time, however, may be obtained
using stochastic quantization techniques, which are not based on a
probability interpretation~\cite{stochquant,Damgaard:1987rr}. In
Refs.~\cite{Berges:2005yt,Berges:2006xc} this has been recently
used to explore the real-time dynamics of an interacting scalar
quantum field theory and $SU(2)$ gauge field theory in $3+1$
dimensions. In real-time stochastic quantization the quantum
ensemble is constructed by a stochastic process in an additional
``Langevin-time'' using the reformulation for the Minkowskian path
integral~\cite{cl,Minkowski}: The quantum fields are defined on a
physical space-time lattice, and the updating employs a Langevin
equation with a complex driving force in an additional, unphysical
``time'' direction. Though more or less formal proofs of
equivalence of the stochastic approach and the path integral
formulation have been given for Minkowskian space-time, not much
is known about the general convergence properties and its
reliability beyond free-field theory or simple
models~\cite{Minkowski,convergence}. Most investigations of
complex Langevin equations concern simulations in Euclidean
space-time with non-real
actions~\cite{Karsch:1985cb,Ambjorn:1986fz}.

In Ref.~\cite{Berges:2006xc} real-time stochastic quantization was
applied to quantum field theory without further optimization. For
$SU(2)$ gauge theory no stable physical solution of the complex
Langevin equation could be observed even for small couplings. The
physical fixed point was found to be approached at intermediate
Langevin-times, however, deviations occurred at later times. The
onset time for deviations could be delayed and physical results
extracted, if the real-time extent of the lattice was chosen to be
sufficiently small on the scale of the inverse temperature. This
procedure provided severe restrictions for actual applications of
the method. In contrast, for self-interacting scalar field theory
stable physical solutions were observed.

In this paper we investigate real-time stochastic quantization for
gauge theories employing an optimized updating procedure for the
Langevin process. We consider optimized updating using stochastic
reweighting or gauge fixing, respectively. These procedures do not
affect the underlying theory but strongly improve the stability
properties of the stochastic dynamics. For $SU(2)$ gauge theory in
$3+1$ dimensions we demonstrate that gauge fixing leads already to
stable physical solution for not too small $\beta \sim 1/g^2$,
where large $\beta$ correspond to going to the continuum limit of
the lattice gauge theory. Where applicable, the results are shown
to accurately reproduce alternative calculations in Euclidean
space-time. In order to gain analytical understanding and to
compare to exact results we also investigate $U(1)$ and $SU(2)$
one-plaquette models.

The paper is organized as follows. In Sec.~\ref{sec:realtime} we
briefly review real-time stochastic quantization for non-Abelian
lattice gauge theory following Ref.~\cite{Berges:2006xc}. The
$U(1)$ one-plaquette model of Sec.~\ref{sec:u1} is used to
introduce the concept of stochastic reweighting in
Sec.~\ref{sec:stochre}. The simplicity of the model allows us to
compare simulation with analytical results and to investigate in
some detail the fixed point structure and convergence properties
is Secs.~\ref{sec:fp} and \ref{sec:stability}. In
Sec.~\ref{sec:su2oneplaquette} we consider the $SU(2)$
one-plaquette model and introduce some important notions that will
be employed for the optimized updating using gauge fixing for the
lattice field theory in Sec.~\ref{sec:su2gaugetheory}. We present
conclusions in Sec.~\ref{sec:conclusions} and an appendix provides
some mathematical details.

\section{Real-time gauge theory}
\label{sec:realtime}

Gauge theories on a lattice are formulated in terms of the
parallel transporter $U_{x,\mu}$ associated with the link from the
neighboring lattice point $x+\hat{\mu}$ to the point $x$ in the
direction of the lattice axis $\mu = 0,1,2,3$. The link variable
$U_{x,\mu}= U^{-1}_{x+ \hat\mu,-\mu}\,$ is an element of the gauge
group $G$. For $G = SU(N)$ or $U(1)$ one has $U_{x,\mu}^{-1} =
U^\dagger_{x,\mu}\,$, however, since we will consider a more
general group space in the context of stochastic quantization this
will not be assumed. Therefore, we keep $U^{-1}_{x,\mu\nu}$ in the
definition of the action, which is described in terms of the gauge
invariant plaquette variable
\begin{equation}
U_{x,\mu\nu} \equiv U_{x,\mu} U_{x+\hat\mu,\nu}
U^{-1}_{x+\hat\nu,\mu} U^{-1}_{x,\nu} \label{eq:plaq}\, ,
\end{equation}
where $U_{x,\nu\mu}^{-1}=U_{x,\mu\nu}\,$. The action on a
real-time lattice reads
\begin{eqnarray}
S[U] &=& - \beta_0 \sum_{x} \sum_i \left\{ \frac{1}{2 {\rm Tr}
{\bf 1}} \left( {\rm Tr}\, U_{x,0i} + {\rm Tr}\, U_{x,0i}^{-1}
\right) - 1 \right\}
\nonumber\\
&& + \beta_s \sum_{x} \sum_{i,j \atop i<j} \left\{ \frac{1}{2 {\rm
Tr} {\bf 1}} \left( {\rm Tr}\, U_{x,ij} + {\rm Tr}\, U_{x,ij}^{-1}
\right) - 1 \right\} \, ,  \label{eq:clgaugeaction}
\end{eqnarray}
with spatial indices $i,j = 1,2,3$. Here the relative sign between
the time-like and the space-like plaquette terms reflects the
Minkowskian metric, and
\begin{equation}
\beta_0 \equiv \frac{2 \gamma {\rm Tr} {\bf 1}}{g_0^2} \,\, ,
\quad \beta_s \equiv \frac{2 {\rm Tr} {\bf 1}}{g_s^2 \gamma} \, ,
\label{eq:ganisoM}
\end{equation}
with the anisotropy parameter $\gamma \equiv a_s/a_t$ on a lattice
of size $(N_s a_s)^3 \times N_t a_t$. Because of the anisotropic
lattice we have introduced the anisotropic bare couplings $g_0$
for the time-like plaquettes and $g_s$ for the space-like
plaquettes.

Using stochastic quantization the real-time quantum configurations
in 3+1 dimensions are constructed by a stochastic process in an
additional (5th)
Langevin-time~\cite{cl,Minkowski,Damgaard:1987rr}. For a
discretization with stepsize $\epsilon$ the Langevin-time after
$n$ steps is $\vartheta_n = n \epsilon$. The discretized Langevin
equation for the link variable reads with the notation
$U^{\prime}_{x, \mu} \equiv U_{x, \mu}(\vartheta_{n+1})$ and
$U_{x, \mu} \equiv U_{x, \mu}(\vartheta_{n})$~\cite{Berges:2006xc}
\begin{eqnarray}
U^{\prime}_{x, \mu} &=& \exp\left\{ i \sum_a \lambda_a
\left(\epsilon\, i D_{x \mu a} S[U] + \sqrt{\epsilon}\, \eta_{x
\mu a} \right)\right\} U_{x, \mu}\, , \label{eq:ALangevinM}
\end{eqnarray}
where differentiation in group space is defined by
\begin{equation}
D_{x \mu a} f(U_{x,\mu}) = \frac{\partial}{\partial \omega}\,
f\left( e^{i \omega \lambda_a} U_{x,\mu}\right)|_{\omega = 0}
\end{equation}
with the generators $\lambda_a$ of the Lie algebra and
$a=1,\ldots,N^2-1$ for $SU(N)$. For the action
(\ref{eq:clgaugeaction}) one has
\begin{eqnarray}
i D_{x \mu a} S[U] &=& - \frac{1}{2N} \sum_{\nu=0 \atop \nu \neq
\mu}^{3} \beta_{\mu\nu} {\rm Tr} \left( \lambda_a U_{x,\mu}
C_{x,\mu\nu}- \bar{C}_{x,\mu\nu} U^{-1}_{x,\mu} \lambda_a \right)
\, , \label{eq:derivS}
\end{eqnarray}
where we have defined $\beta_{ij} \equiv \beta_s$, $\beta_{0i}
\equiv \beta_{i0} \equiv - \beta_0$ and
\begin{eqnarray}
C_{x,\mu\nu} &=& U_{x+\hat\mu,\nu} U^{-1}_{x+\hat\nu,\mu}
U^{-1}_{x,\nu} + U^{-1}_{x+\hat\mu-\hat\nu,\nu}
U^{-1}_{x-\hat\nu,\mu} U_{x-\hat\nu,\nu} \, ,
\nonumber\\
\bar{C}_{x,\mu\nu} &=&
U_{x,\nu}U_{x+\hat\nu,\mu}U^{-1}_{x+\hat\mu,\nu} +
U^{-1}_{x-\hat\nu,\nu} U_{x-\hat\nu,\mu}U_{x+\hat\mu-\hat\nu,\nu}
\, .
\end{eqnarray}
With $U_{x,\mu} C_{x,\mu\nu} =  U_{x,\mu\nu} + U_{x,\mu(-\nu)}$
and $\bar{C}_{x,\mu\nu} U^{-1}_{x,\mu} = U_{x,\mu\nu}^{-1} +
U_{x,\mu (-\nu)}^{-1}$ one observes that the sum in
Eq.~(\ref{eq:derivS}) is over all possible plaquettes containing
$U_{x,\mu}\,$. Following Ref.~\cite{Berges:2006xc} the Gaussian
noise $\eta_{x \mu a} \equiv \eta_{x \mu a}(\vartheta_n)$
appearing in (\ref{eq:ALangevinM}) is taken to be real and
satisfies\footnote{It was suggested in earlier
literature~\cite{Damgaard:1987rr} to replace $\delta_{\mu\nu}$ on
the right-hand side of Eq.~(\ref{eq:realtimenoise}) by
$g_{\mu\nu}$. However, in this case solutions of the Langevin
evolution would not respect the Dyson-Schwinger identities of the
underlying quantum field theory, as is shown in
Ref.~\cite{Berges:2006xc}.}
\begin{equation}
\langle \eta_{x \mu a} \rangle = 0 \,, \qquad \langle \eta_{x \mu
a}\, \eta_{y \nu b} \rangle = 2\, \delta_{\mu\nu} \delta_{xy}
\delta_{ab} \, . \label{eq:realtimenoise}
\end{equation}
Expectation values for observables can be obtained from solving
equation (\ref{eq:ALangevinM}) for sufficiently large
Langevin-time by performing noise averages or, alternatively, from
Langevin-time averages~\cite{montvay}.

For instance, specifying to $SU(2)$ gauge theory
$\lambda_a=\sigma_a$ ($a=1,2,3$) represent the Pauli matrices, and
one can make further simplifications using $ {\rm Tr} (U^{-1}
\sigma^a) = -{\rm Tr} ( U \sigma^a)$ for any element $U \in
SU(2)$. The latter simplification also holds for $U \in SL(2,{\bf
C})$. This is relevant since possible solutions of
Eq.~(\ref{eq:ALangevinM}) may respect this enlarged symmetry
group. Taking \beq U_{x,\mu} \equiv e^{i A_{x \mu a} \sigma_a/2}
\eeq the vector fields $A_{x \mu a}$ need not to be real for $U
\in SL(2,{\bf C})$. The complex matrix $A_{x \mu}^a \sigma_a$
still remains traceless, however, the Hermiticity properties are
lost. As a consequence, it is no longer possible to identify
$U^{\dagger}$ with $U^{-1}$ as is taken into account in
Eq.~(\ref{eq:clgaugeaction}). This corresponds to an extension of
the original $SU(2)$ manifold to $SL(2,{\bf C})$ for the Langevin
dynamics. Only after taking noise or Langevin-time averages,
respectively, the expectation values of the original $SU(2)$ gauge
theory are to be recovered. Accordingly, if the Langevin flow
converges to a fixed point solution of Eq.~(\ref{eq:ALangevinM})
it automatically fulfills the infinite hierarchy of
Dyson-Schwinger identities of the original
theory~\cite{Berges:2006xc,xue}.

\section{Optimized updating: simple examples}

\subsection{One-plaquette model with $U(1)$ symmetry}
\label{sec:u1}

\subsubsection{Direct integration}

As a first example we consider the one-plaquette model with $U(1)$
symmetry. For $U= e^{i \varphi}$ the action is given by
\begin{eqnarray}
S_0 \,=\,  \frac{\beta}{2} \left( U + U^{-1} \right) \,=\, \beta
\cos \varphi \,  \label{eq:ac0}
\end{eqnarray}
with real coupling parameter $\beta$. The one-plaquette "partition
function" is
\begin{eqnarray}
Z_0 \,=\, \int_{0}^{2\pi}{\rm d}\varphi \, \e^{i S_0} \, = \, 2
\pi J_0(\beta) \, ,\label{eq:u1int}
\end{eqnarray}
where $J_0(\beta)$ denotes a Bessel function of the first
kind~\cite{Grad}, with \mbox{$J_0(1) \simeq 0.765$}. The average
of an observable $O(\varphi)$ is obtained as
\begin{eqnarray}
\langle O \rangle_0 \,=\, \frac{1}{Z_0} \int_{0}^{2\pi}{\rm
d}\varphi \, \e^{i S_0}\, O(\varphi) \, .\label{eq:avO}
\end{eqnarray}
For real $\beta$ the integrand in Eq.\ (\ref{eq:avO}) is not
positive definite, which mimics certain aspects of more
complicated theories in Minkowskian space-time that will be
considered below. In contrast to those more realistic theories,
the one-plaquette model has the advantage that the elementary
integrals can be performed and the results directly compared to
those obtained from stochastic methods.

\subsubsection{Complex Langevin equation}
\label{sec:complex}

In principle, adding a Langevin-time dependence $\varphi \to
\varphi(\vartheta_n)$ all observables can be computed from a
solution of the discretized Langevin equation
\begin{eqnarray}
\varphi' &=& \varphi + i \epsilon \, \frac{\partial
S_0(\varphi)}{\partial \varphi} + \sqrt{\epsilon}\, \eta \nonumber\\
&=& \varphi - i\epsilon \, \beta \sin \varphi + \sqrt{\epsilon}\,
\eta \, , \label{eq:clange}
\end{eqnarray}
using a notation as in Eq.\ (\ref{eq:ALangevinM}). The real
Gaussian noise fulfills
\begin{equation}
\langle \eta \rangle \,=\, 0 \quad , \qquad \langle \eta\eta
\rangle \,=\, 2 \label{eq:noiseu1}
\end{equation}
according to Eq.~(\ref{eq:realtimenoise}).

In view of the aim to compute plaquette averages, we consider the
average of the function $e^{i l \varphi}$ with integer
$l$.\footnote{We will not investigate here the question of
defining roots of group elements from stochastic processes and do
not consider non-integer $l$.\label{foot:f1}} We first compute
averages by analytic or direct numerical integration according to
Eq.\ (\ref{eq:avO}), and then compare to the result for the same
quantity obtained from a stochastic process using Eq.\
(\ref{eq:clange}). For $\beta =1$ and $l=1$ one obtains from Eq.\
(\ref{eq:avO})
\begin{equation}
\langle e^{i \varphi} \rangle_0 \, =\, i\, \frac{J_1(1)}{J_0(1)}
\, \simeq\, i\, 0.575\, , \label{eq:ex}
\end{equation}
which can be compared to the result from the solution of the
Langevin equation (\ref{eq:clange}):
\begin{equation}
\langle e^{i \varphi} \rangle_0 \, \stackrel{\rm without \atop
optimization}{=}\, -0.009(\pm 0.006) - i\, 0.00006(\pm 0.00007) \,
.
\end{equation}
This result was obtained for $\beta=1$ using a Langevin stepsize
$\epsilon = 10^{-5}$ from a Langevin-time average over $10^{10}$
steps. The error in brackets gives the statistical fluctuation of
the average. One observes that the simulation yields a wrong
result that is compatible with zero, in contrast to the
non-vanishing imaginary value (\ref{eq:ex}) obtained analytically.
A similar failure of the stochastic method can be observed for
averages of other functions as well.

\subsubsection{Optimized updating by stochastic reweighting}
\label{sec:stochre}

The very same average values as above may be computed with the
help of the partition function for a different action
$S_\alpha\,$,
\begin{eqnarray}
Z_\alpha &=& \int_{0}^{2\pi}{\rm d}\varphi \, \e^{i S_\alpha} \,
,\label{eq:pu1int}
\end{eqnarray}
from
\begin{eqnarray}
\langle O \rangle_\alpha &=& \frac{1}{Z_\alpha}
\int_{0}^{2\pi}{\rm d}\varphi \, \e^{i S_\alpha} O(\varphi) \,
\label{eq:avp}
\end{eqnarray}
using standard reweighting techniques. If we define
\begin{eqnarray}
\omega_\alpha = e^{i(S_0-S_\alpha)} \label{eq:om}
\end{eqnarray}
then the expectation value (\ref{eq:avO}) can be identically
written as
\begin{eqnarray}
\langle O \rangle_0 \,=\, \frac{ \int_{0}^{2\pi}{\rm d}\varphi \,
\e^{i S_\alpha}\, \omega_\alpha\, O(\varphi)}{\int_{0}^{2\pi}{\rm
d}\varphi \, \e^{iS_\alpha} \omega_\alpha} \,=\, \frac{\langle
\omega_\alpha\, O\, \rangle_\alpha}{\langle
\omega_\alpha\rangle_\alpha} \, .\label{eq:ow}
\end{eqnarray}
We will consider the family of actions
\begin{eqnarray}
S_\alpha \,=\, S_0 + \alpha \varphi \,=\, \beta \cos \varphi +
\alpha \varphi  \label{eq:acaplha}
\end{eqnarray}
with integer$^{\ref{foot:f1}}$ $\alpha$, such that the reweighting
function (\ref{eq:om}) reads
\begin{equation}
\omega_\alpha = e^{- i \alpha \varphi} \, .
\end{equation}
For the action (\ref{eq:acaplha}) the discretized Langevin
equation is given by
\begin{eqnarray}
\varphi' &=& \varphi - i\epsilon \, \beta \sin \varphi + i\epsilon
\, \alpha + \sqrt{\epsilon}\, \eta \, . \label{eq:clangedis}
\end{eqnarray}
As a consequence, for $\alpha \not = 0$ the average value in Eq.\
(\ref{eq:ow}) is computed with the help of a different stochastic
process than in Sec.\ \ref{sec:u1}. For $\beta = 1$ and the very
same Langevin parameters as above we obtain for $\alpha = 1$ the
simulation result:
\begin{equation}
\langle e^{i \varphi} \rangle_0 \, =\, \frac{ \langle 1
\rangle_{\alpha=1}}{\langle e^{-i \varphi} \rangle_{\alpha=1}} \,
= \, -0.001(\pm 0.003) + i\, 0.575(\pm 0.0004)\, ,
\end{equation}
which is close to the exact result given in Eq.~(\ref{eq:ex}), in
contrast to the failure of the method without reweighting.

\begin{table}
\begin{center}
\begin{tabular}{|c|c||c|c||c|c|}
\hline
$\alpha$ & $l$ & exact Re &   stochastic Re & exact Im & stochastic Im  \\
\hline\hline
$0$ & $1$ & $0$ & $ -0.00875(\pm 0.006) $ & $0.575$ & $ 5.88\!\cdot\! 10^{-5}(\pm 7\!\cdot\! 10^{-5}) $ \\
$0$ & $-1$ & $0$ & $ -0.00218(\pm 0.006) $ & $0.575$ & $ 8.3\!\cdot\! 10^{-5}(\pm 8\!\cdot\! 10^{-5}) $ \\
\hline
$1$ & $1$ & $0$ & $\! -0.000626 (\pm 0.0007) \!\!$ & $0.261$ & $ 0.261 (\pm 0.0005) $ \\
$1$ & $-1$ & $0$ & $ -0.00292 (\pm 0.007) $ & $-1.74$ & $ -1.74 (\pm 0.0005) $ \\
$1$ & $2$ & $\!-0.0445\!$ & $ -0.0442 (\pm 0.0002) $ & $0$ & $\! -0.000221 (\pm 0.0003) \!$ \\
$1$ & $-2$ & $1$ & $ 0.998 (\pm 0.03) $ & $0$ & $ 0.012 (\pm 0.03) $ \\
\hline
$2$ & $1$ & $0$ & $\! -0.000192(\pm 0.0003) \!\!$ & $0.17$ & $ 0.17(\pm 0.0002)$ \\
$2$ & $-1$ & $0$ & $ -0.00788(\pm 0.009) $ & $-3.83$ & $ -3.83(\pm 0.0002)$ \\
$2$ & $2$ & $\!-0.0216\!$ & $ -0.0216(\pm 6\!\cdot\! 10^{-5})$ & $0$ & $\! -3.5\!\cdot\! 10^{-5}(\pm 7\!\cdot\! 10^{-5})\!\!$\\
$2$ & $-2$ & $-6.66$ & $ -6.66(\pm 0.03)$ & $0$ & $ 0.0338(\pm 0.06)$ \\
\hline
\end{tabular}
\end{center}
\caption{\small Averages $\langle e^{i l \varphi} \rangle_\alpha$
for different values of the reweighting parameter $\alpha$ and
{\em fixed} $\beta = 1$. Listed are results for the real and
imaginary part of $\langle e^{i l \varphi} \rangle_\alpha$ denoted
as Re and Im, respectively. Compared are exact results from direct
integration with simulation results using a stochastic process.
For the exact values three significant digits are given. For the
stochastic method the given error in brackets reflects statistical
fluctuations.} \label{tab:tabu2}
\end{table}
In Table \ref{tab:tabu2} we list results for averages using
various values of $\alpha$ and {\em fixed} $\beta =1$. Shown are
the results for Re$\langle e^{i l \varphi}\rangle_\alpha$ and
Im$\langle e^{i l \varphi}\rangle_\alpha$ both from direct
integration ("exact") of Eq.~(\ref{eq:avp}) and using a stochastic
process ("stochastic") according to Eq.~(\ref{eq:clangedis}),
where we use the same Langevin parameters for the numerics as
before. The first two rows correspond to a reweighting parameter
$\alpha = 0$, i.e.\ no reweighting, showing the strong
disagreement of simulation and exact results in this case. We
obtain similarly bad results as long as $\alpha \lesssim \beta$.
In contrast, Table \ref{tab:tabu2} shows for $\alpha = 1$ accurate
values obtained from simulation. We note that in this case the
value for $\beta$ agrees with the value chosen for the reweighting
parameter $\alpha$, i.e.\ $\alpha = \beta = 1$. For $\alpha = 2$
the results are still accurate, which we observe also for several
other $\alpha$ and $\beta$ as long as $\alpha \gtrsim \beta$. We
typically need to collect substantially more statistics for
negative $l$ to keep statistical errors small as compared to the
case with positive $l$, which will be addressed in
Sec.~\ref{sec:stability}.

\begin{figure}[t]
\begin{center}
\epsfig{file=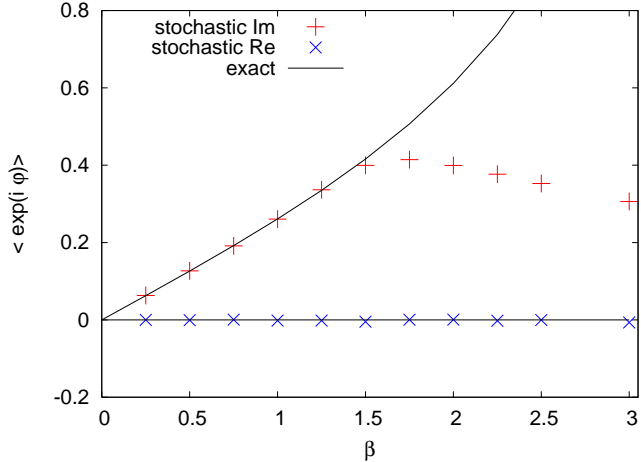,width=9.cm}
\end{center}
\vspace*{-0.5cm} \caption{\small The real and imaginary part of
the average $\langle e^{i \varphi} \rangle_{\alpha=1}$ as a
function of $\beta$. The lines represent averages obtained from
direct integration, while the symbols are measurements using a
stochastic process.} \label{fig:betadep}
\end{figure}
\begin{figure}[t]
\begin{center}
\epsfig{file=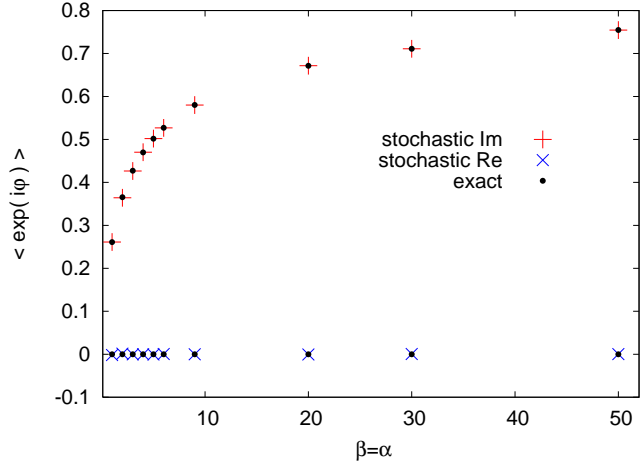,width=9.cm}
\end{center}
\vspace*{-0.5cm} \caption{Shown are the real and the imaginary
part of the average $\langle e^{i \varphi} \rangle$ as a function
of integer $\alpha$ with $\beta = \alpha$.} \label{fig:expiphi}
\end{figure}

The dependence of the accuracy of the simulation outcome on the
relative size of $\alpha$ and $\beta$ is further illustrated in
Fig.\ \ref{fig:betadep}, where we consider results for various
values of $\beta$ and fixed $\alpha=1$ with $l=1$. Shown are the
values for the averages Re$\langle e^{i
\varphi}\rangle_{\alpha=1}$ and Im$\langle e^{i
\varphi}\rangle_{\alpha=1}$ again from direct integration (lines),
which yields
\begin{eqnarray}
\frac{1}{Z_1}\int_0^{2 \pi} {\rm d}\varphi\, e^{i(\beta \cos
\varphi + \varphi)} e^{i \varphi} &=& i\,
\frac{J_2(\beta)}{J_1(\beta)} \, , \label{eq:avu1}
\end{eqnarray}
as well as using a stochastic process (symbols). One observes that
for $\beta \lesssim \alpha$ the simulation method gives accurate
results for this quantity, while for somewhat larger $\beta$ they
can deviate substantially. In Fig.~\ref{fig:expiphi} we show
results for $\langle e^{i \varphi} \rangle$ as a function of
integer values of $\alpha$ for the case $\beta = \alpha$, which
shows very good agreement between stochastic averages and direct
integrations.$^{\ref{foot:f1}}$

The above analysis shows that the accuracy of the simulation
method depends strongly on the employed stochastic process. It
suggests that for a given model there is an optimized stochastic
process, which may yield accurate results. The conditions that
have to be met in order to obtain quantitative estimates will be
further explained in the following.

\subsubsection{Fixed point structure}
\label{sec:fp}

The stationary solutions of the noise averaged Langevin equation
(\ref{eq:clangedis}) are determined by
\begin{equation}
\label{eq:fixp}
 -\beta \langle \sin \varphi \rangle_\alpha + \alpha = 0 \, .
 \label{eq:fpc1}
\end{equation}
Taking into account that the dynamical variable can become complex
with $\varphi = \varphi_R + i \varphi_I$ this is equivalent to the
statement that the equations
\begin{eqnarray}
\langle \cos \varphi_R\, \sinh \varphi_I \rangle_\alpha & = & 0 \,
, \nonumber\\
\langle \sin \varphi_R\, \cosh \varphi_I \rangle_\alpha &=&
\frac\alpha\beta \, , \label{eq:fpc}
\end{eqnarray}
have to be simultaneously satisfied. Neglecting fluctuations,
i.e.\ disregarding for a moment the noise averages,
Eqs.~(\ref{eq:fpc}) correspond to the classical fixed point
condition $\partial S/\partial \varphi\, (\varphi = \varphi^*) =
0$. The first of these equations constrains either $\varphi_I =
\varphi_I^* = 0$ or $\varphi_R = \varphi_R^* = \pi/2$ or $3\pi/2$
for the real part of the fixed point value. Taking into account
the second equation of (\ref{eq:fpc}), there are two distinct real
solutions for $\beta > |\alpha|$. There are two complex conjugated
solutions for $\beta < |\alpha|$ and for $\beta = |\alpha|$ there
is a single real solution. We emphasize that these last statements
about the possible fixed points just consider the first derivative
of the {\em classical} action with respect to the dynamical
variable, instead of the noise averaged quantity $\langle
\partial S/\partial \varphi \rangle$.

The classical approximation becomes exact for $\beta = |\alpha|
\to \infty$. In this limit the integral (\ref{eq:ow}) is dominated
by the value where the oscillatory integrand shows slowest
variation in $\varphi$, i.e.\ for $\varphi = \varphi^*$. As a
consequence one obtains
\begin{eqnarray}
\lim\limits_{\beta \to \infty} \frac{ \int_{0}^{2\pi}{\rm
d}\varphi \, \e^{i \beta (\cos \varphi \pm \varphi)}\,
O(\varphi)}{\int_{0}^{2\pi}{\rm d}\varphi \, \e^{i \beta (\cos
\varphi \pm \varphi)}} \,=\, O(\varphi^*) \, , \label{eq:limitfp}
\end{eqnarray}
with $\varphi^* = \pi/2$ for the positive sign in the exponent of
the integrand, i.e.\ $\alpha=\beta
> 0$, and $\varphi^* = 3\pi/2$ for the negative sign corresponding to
$\alpha=-\beta < 0$. We will see below that important aspects of
the results for finite $\beta$ of Sec.~\ref{sec:stochre} can be
understood already from the classical fixed point condition, i.e.\
neglecting fluctuations.

\begin{figure}[t]
\begin{center}
\epsfig{file=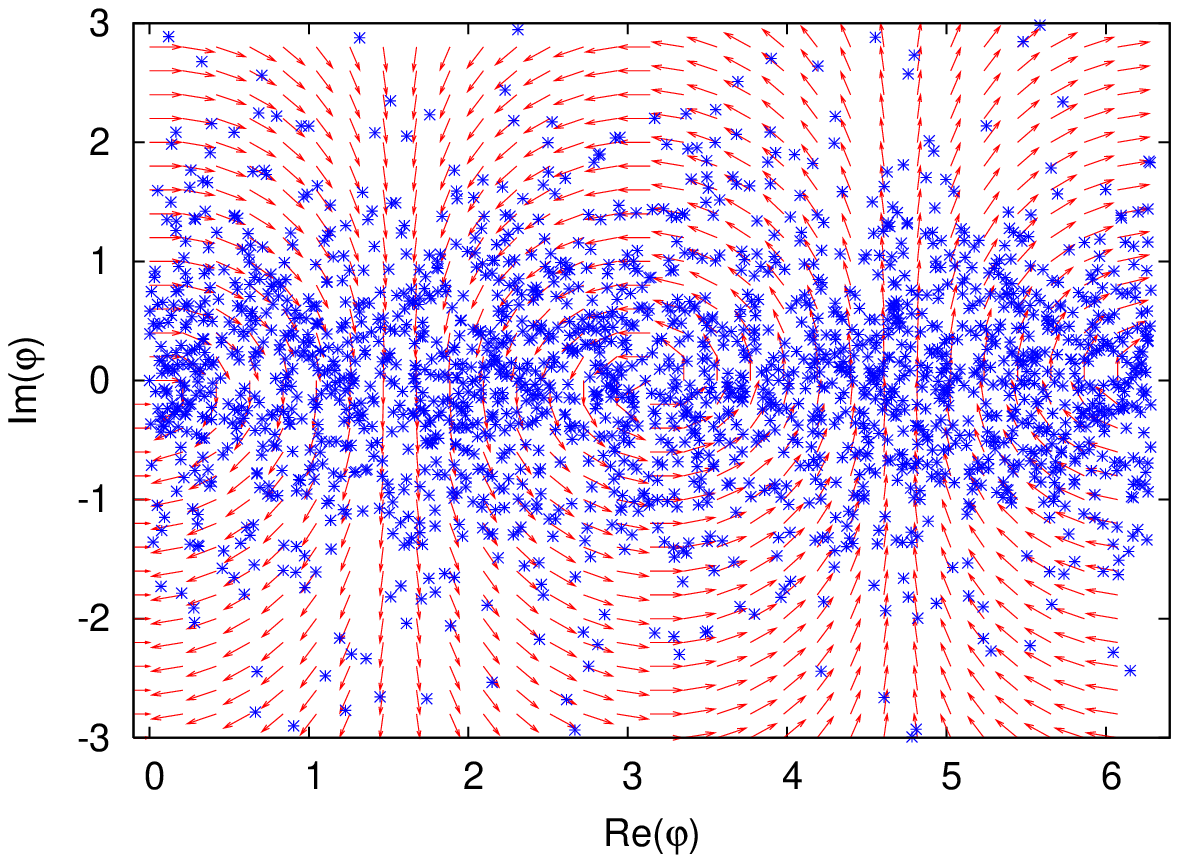,width=6.7cm}
\epsfig{file=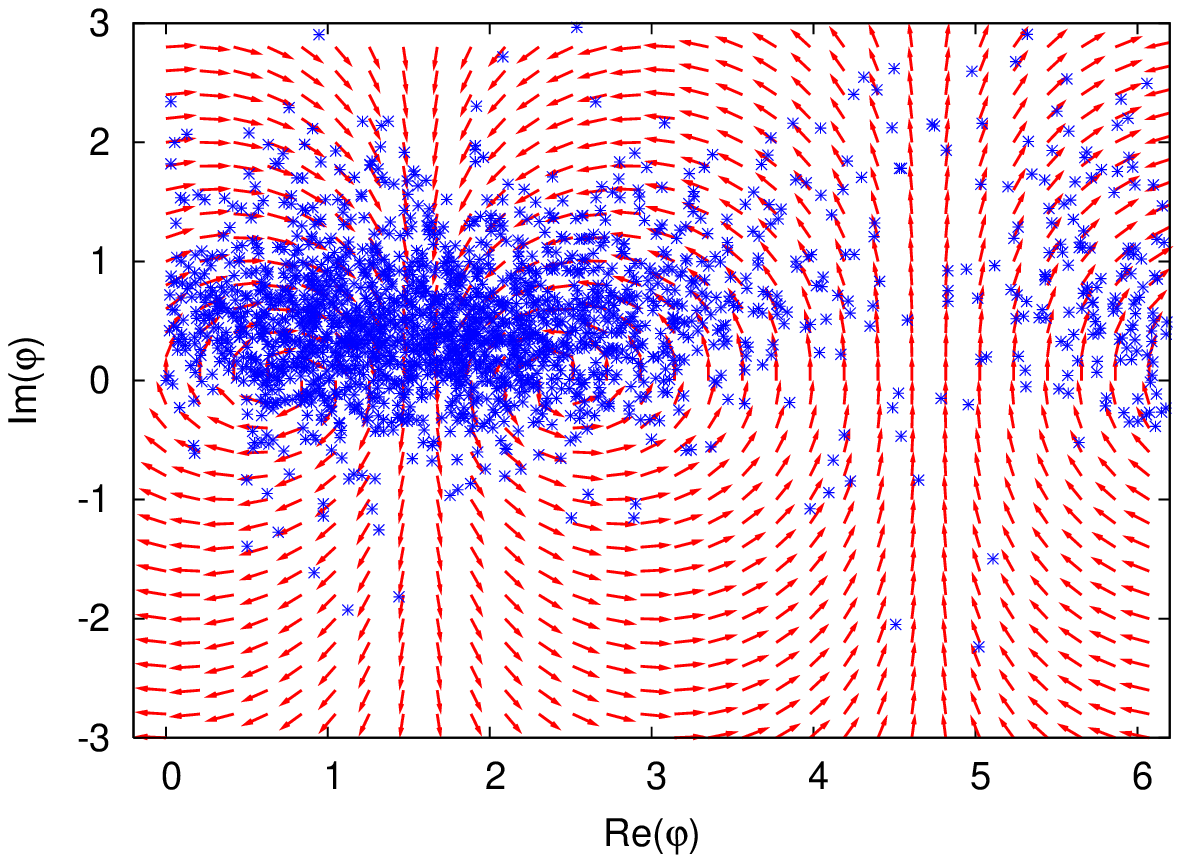,width=6.7cm}
\end{center}
\vspace*{-0.5cm} \caption{\small Shown is the real and imaginary
part of $\partial S/\partial \varphi$ plotted as a vector with
origin at each $\varphi$-value and with normalized length for
better visibility. Here we consider $\beta \gtrsim \alpha$, for
which the stochastic method fails. For the left graph \mbox{$1 =
\beta > \alpha=0$}, while the right graph employs \mbox{$1.5 =
\beta > \alpha = 1$}. From the arrows one can infer the two
non-attractive, classical fixed points on the real axis. Also
shown is the distribution of $\varphi$ as obtained from the full
solution of the respective complex Langevin equation. One observes
relatively wide distributions with values having positive as well
as negative imaginary parts. This has to be compared to
Fig.~\ref{fig:drift2} below for the case $\beta \lesssim \alpha$.}
\label{fig:drift1}
\end{figure}
For Fig.~\ref{fig:drift1}, we have evaluated $\partial S/\partial
\varphi$ for the range of values $0 \le \varphi_R \le 2 \pi$ and
$-\pi \le \varphi_I \le \pi$ and plotted its real and imaginary
part as a vector with origin at each $\varphi$-value. The size of
the real and imaginary part of $\partial S/\partial \varphi$
determines the direction and angle of each vector, however, we
normalized their length for better visibility. The left figure
employs $\alpha = 0$, for which one infers from
Eqs.~(\ref{eq:fpc}) two classical fixed points at $\varphi^*=0$
and $\varphi^*= \pi$. From the arrows it can be seen that these do
not correspond to attractive fixed points, where the drift term in
the Langevin equation (\ref{eq:clange}) would tend to focus the
Langevin flow. Instead they are "circular" with opposite rotation
directions for the two points. Taking into account the
$2\pi$-periodicity of the dynamical variable one observes that the
fixed points are equidistantly separated along the real axis in
this case.

These properties of the classical fixed point determine to a large
extend the full Langevin flow, i.e.\ the behavior of the dynamical
variable in the presence of fluctuations due to the noise term
$\sim \eta$ in Eq.~(\ref{eq:clangedis}). In order to visualize the
distribution of $\varphi$, we make snapshots of the Langevin
process with a time-step of $ \Delta \vartheta= 1$, and plot the
values in the complex plane. The resulting distribution for
\mbox{$1 = \beta > \alpha=0$}, i.e.\ without reweighting, is given
in the left graph of Fig.~\ref{fig:drift1}. One observes that the
values are rather evenly distributed along the real axis and very
loosely centered around it in the complex plane.

For comparison the right graph of Fig.~\ref{fig:drift1} shows the
corresponding results for \mbox{$1.5 = \beta > \alpha = 1$}. One
observes the two circular fixed points, which are, however, no
longer equidistantly distributed along the real axis. Accordingly,
the distribution obtained from the full Langevin dynamics varies
considerably along the real axis, with a larger density of points
where the classical fixed points are closest to each other. The
larger value for $\beta$ compared to the one employed for the left
graph leads to a smaller width of the distribution in the complex
plane. However, in contrast to the case $\alpha = 0$, one observes
that the values for the reweighted theory are predominantly
localized in the positive-$\varphi_I$ half-plane.

\begin{figure}[t]
\begin{center}
\epsfig{file=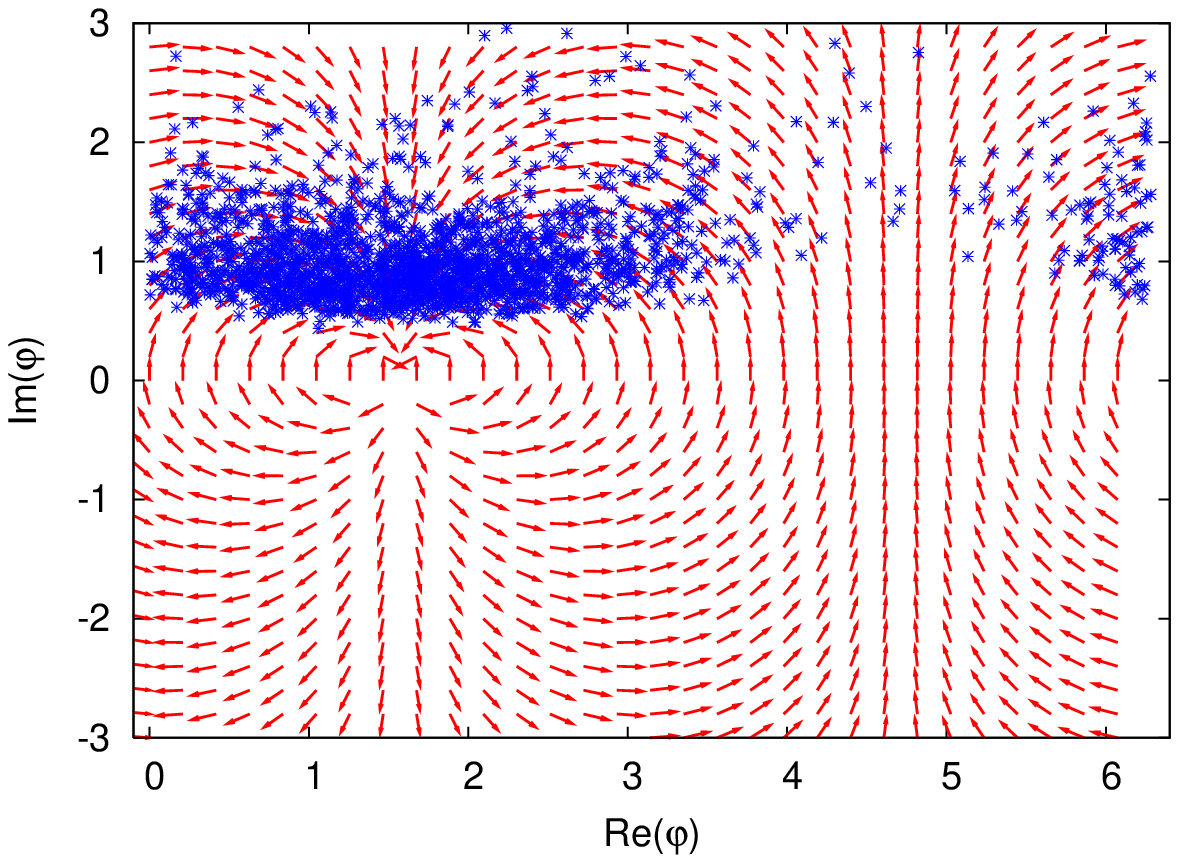,width=6.7cm}
\epsfig{file=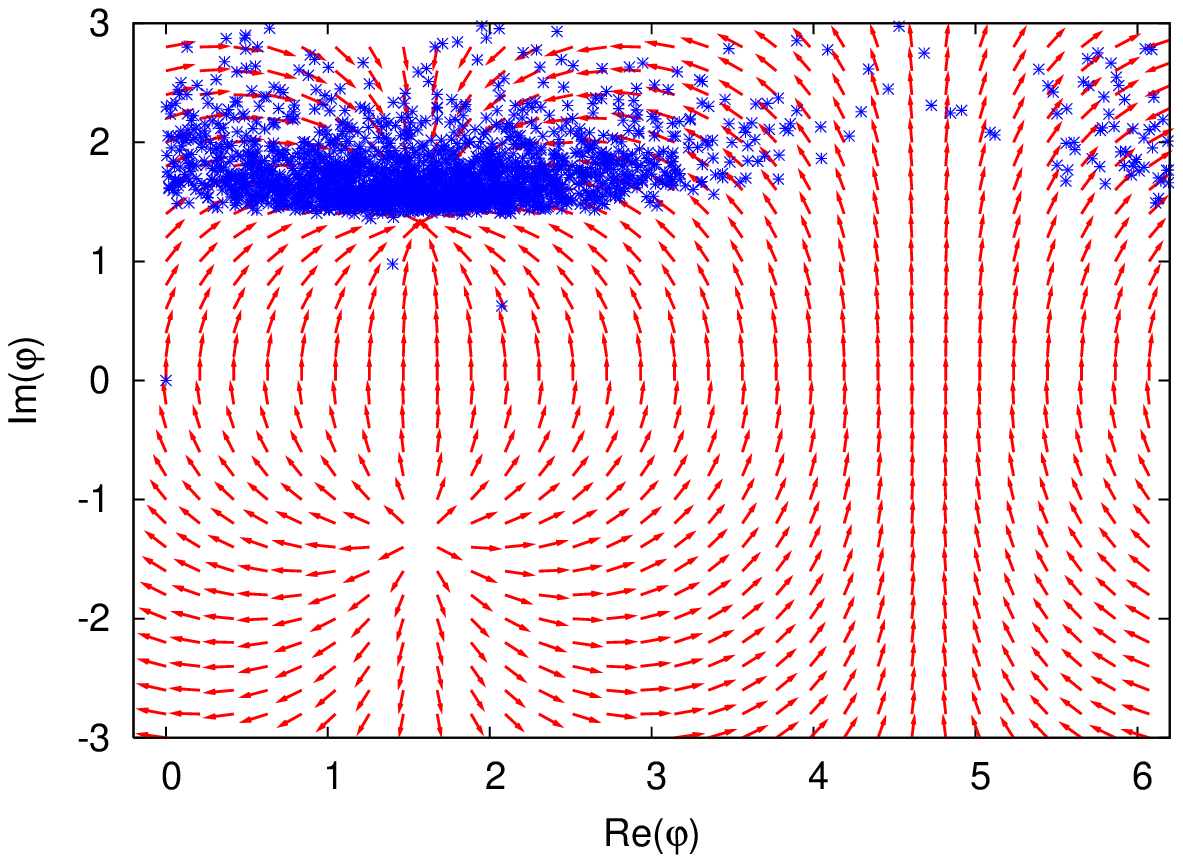,width=6.7cm}
\end{center}
\vspace*{-0.5cm} \caption{\small Same as in Fig.~\ref{fig:drift1}
but for $\beta = \alpha = 1$ (left) and $0.5 = \beta < \alpha = 1$
(right). For $\beta = \alpha$ one observes the real fixed point
with all arrows pointing towards it for positive imaginary part
and away from it for negative imaginary part of the dynamical
variable. For $\beta < \alpha$ an attractive fixed point in the
positive-$\varphi_I$ half-plane and a repulsive one in the lower
half-plane appears. The comparably narrow distribution for
$\varphi$ shows that the Langevin flow spends practically all the
time near the attractive side with positive imaginary parts. For
the considered cases, where $\beta \lesssim \alpha$, the
stochastic method is found to give quantitative results.}
\label{fig:drift2}
\end{figure}
The latter tendency continues when $\beta$ is decreased with
respect to $|\alpha|$. The left graph of Fig.~\ref{fig:drift2}
shows results for $\beta = \alpha = 1$, for which only one
classical fixed point at a real value appears. This stationary
point is neither attractive nor repulsive, with all arrows
pointing towards the fixed point for positive imaginary part of
$\varphi$ and away from it for negative imaginary part. From the
distribution in the left graph of Fig.~\ref{fig:drift2} one
observes that the dynamical variable spends most of the
Langevin-time near the attractive side of the fixed point.
Finally, for $0.5 = \beta < \alpha = 1$ the two complex fixed
points are visible from the right graph of Fig.~\ref{fig:drift2}.
The one in the positive-$\varphi_I$ half-plane is attractive with
all arrows pointing towards the fixed point, while the other in
the negative-$\varphi_I$ half-plane is repulsive. Accordingly, the
distribution shown in the right graph of Fig.~\ref{fig:drift2} is
practically entirely localized in the positive-$\varphi_I$
half-plane.

\begin{figure}[t]
\begin{center}
\epsfig{file=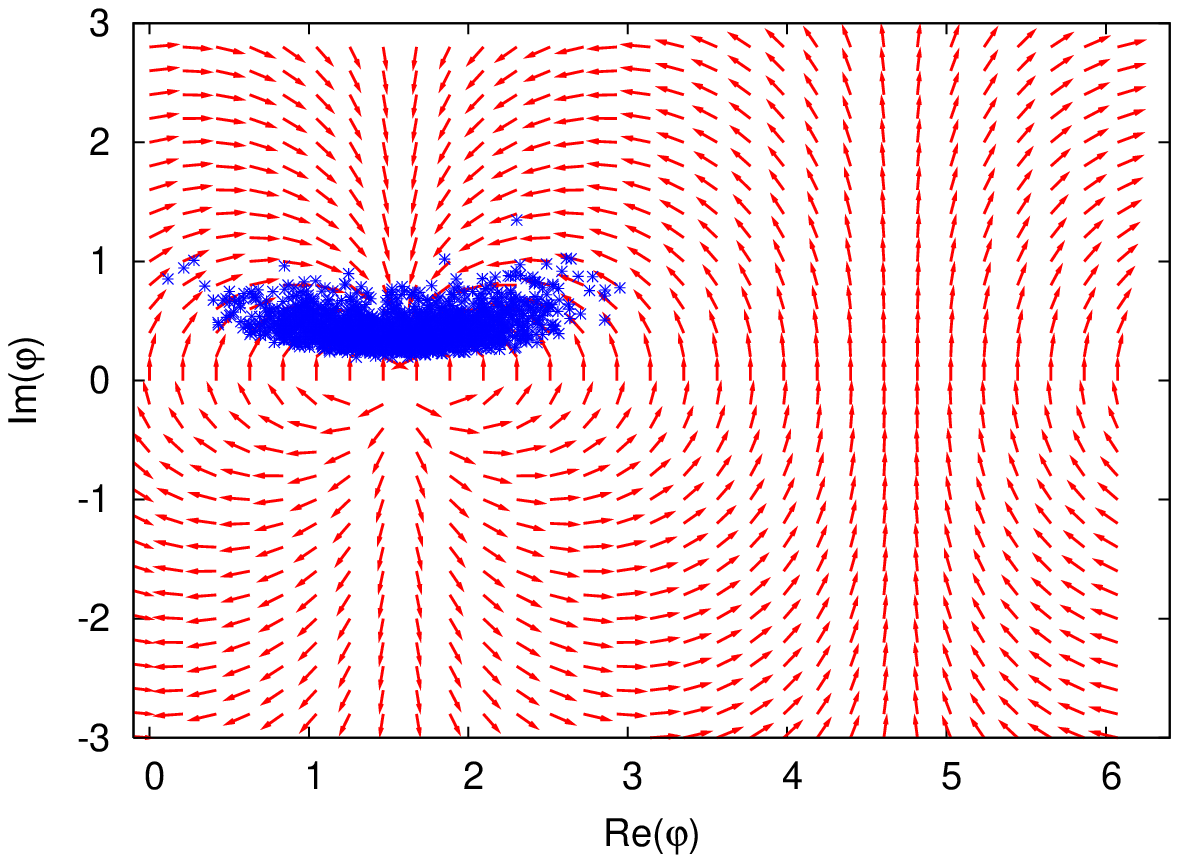,width=6.7cm}
\epsfig{file=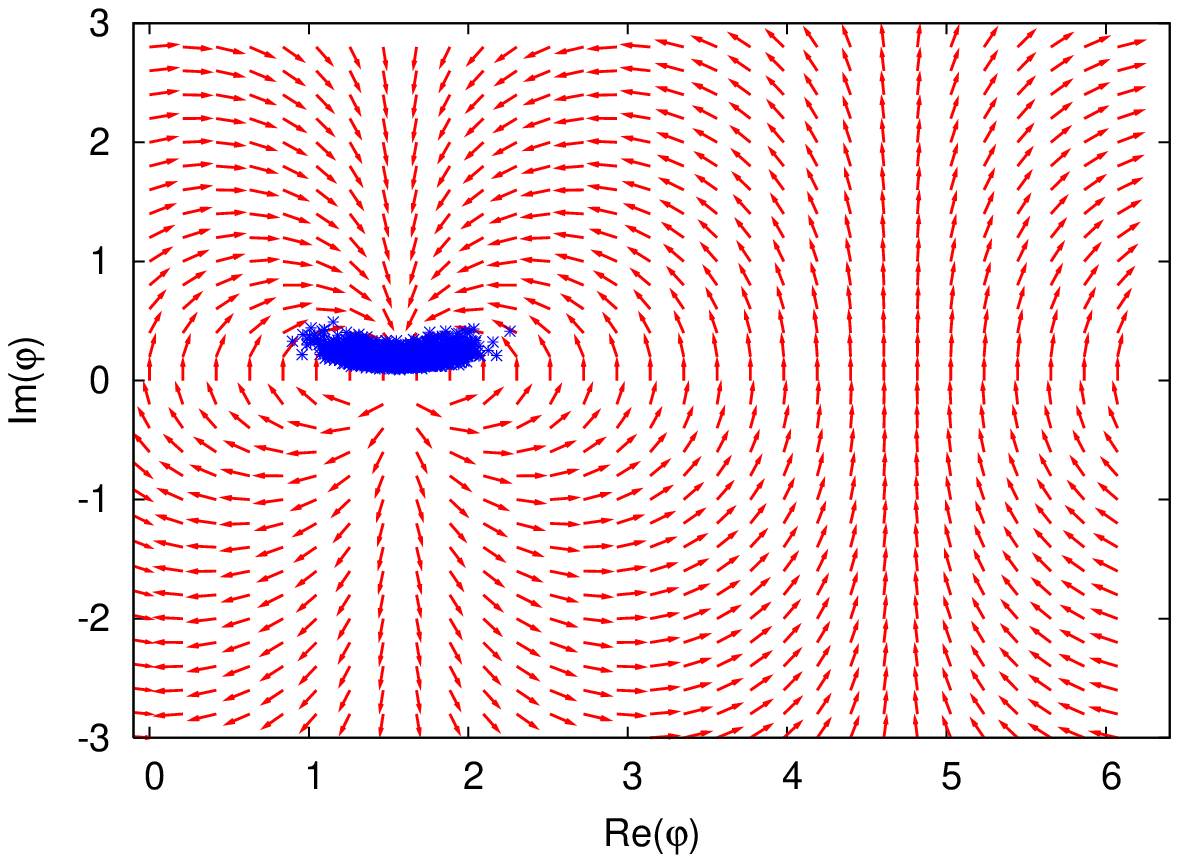,width=6.7cm}
\end{center}
\vspace*{-0.5cm} \caption{\small Same as for the left graph of
Fig.~\ref{fig:drift2} but for $\beta = \alpha = 10$ (left) and
$\beta = \alpha = 100$ (right).} \label{fig:width}
\end{figure}
Fig.~\ref{fig:width} shows the same as the left graph of
Fig.~\ref{fig:drift2} but with larger $\beta = \alpha = 10$ (left)
and $\beta = \alpha = 100$ (right). As the value for $\beta$ is
increased the distribution is more and more centered near the
vicinity of the classical fixed point. This reflects the fact that
the ($\alpha=\beta$)-reweighted one-plaquette model has a well
defined limit $\beta \to \infty$ described by
Eq.~(\ref{eq:limitfp}), in which fluctuations are suppressed. This
qualitative property of a suppression of fluctuations for large
$\beta$ will also be encountered in the discussion for the
non-Abelian field theory in Sec.~\ref{sec:su2gaugetheory}.

\subsubsection{Convergence}
\label{sec:stability}

In view of the above findings we consider in the following
analytical arguments under which conditions convergence of the
stochastic process to accurate results may be expected. For this
we associate the stochastic process (\ref{eq:clangedis}) to a
Langevin-time dependent distribution
$P_\alpha(\varphi;\vartheta_n)$ for the stochastic variable
$\varphi$. Using a notation as in Eq.~(\ref{eq:clangedis}) by
writing $P^\prime_\alpha(\varphi) \equiv
P_\alpha(\varphi;\vartheta_{n+1})$ and $P_\alpha(\varphi) \equiv
P_\alpha(\varphi;\vartheta_{n})$ its evolution can be obtained
from
\begin{eqnarray}
P^\prime_\alpha(\varphi^\prime) = \Bigg\langle \int {\rm
d}\varphi\, P_\alpha(\varphi)\, \delta\left(\varphi^\prime -
\varphi - i \epsilon\, \frac{\partial S_\alpha(\varphi)}{\partial
\varphi} - \sqrt{\epsilon}\, \eta \right) \Bigg\rangle  ,
\end{eqnarray}
where the brackets indicate noise average according to
Eq.~(\ref{eq:noiseu1}). Expanding the $\delta$-functions and
keeping only terms up to order $\epsilon$ gives the Fokker-Planck
equation
\begin{eqnarray}
\frac{1}{\epsilon} \left( P^\prime_\alpha - P_\alpha
\right)(\varphi) &=& \frac{\partial}{\partial \varphi}
\left(\frac{\partial P_\alpha}{\partial \varphi} - i P_\alpha\,
\frac{\partial S_\alpha}{\partial \varphi} \right)(\varphi) +
{\cal O}(\epsilon)\, . \label{eq:fp}
\end{eqnarray}
In the continuum limit $\epsilon \to 0$ we write
\begin{eqnarray}
\frac{\partial P_\alpha(\varphi;\vartheta)}{\partial \vartheta}
&=& - H_{FP}(\varphi) P_\alpha(\varphi;\vartheta)\, ,
\label{eq:contfp}
\end{eqnarray}
with the Fokker-Planck "Hamiltonian"
\begin{eqnarray}
H_{FP}(\varphi) &=& - \frac{\partial^2}{\partial \varphi^2} + i\,
\frac{\partial S_\alpha(\varphi)}{\partial
\varphi}\frac{\partial}{\partial \varphi}  + i\, \frac{\partial^2
S_\alpha(\varphi)}{\partial \varphi^2}\, .
\end{eqnarray}

We first consider the limit of large $\beta = |\alpha|$, where we
have seen in Sec.~\ref{sec:fp} that the stochastic process
converges and is properly governed by the classical fixed point
$\partial S/\partial \varphi\, (\varphi = \varphi^*) = 0$.
According to Eq.~(\ref{eq:limitfp}) the stationary distribution
\begin{eqnarray}
\lim\limits_{\vartheta \to \infty} P_\alpha(\varphi;\vartheta) =
P_\alpha^*(\varphi)
\end{eqnarray}
in this case is described by
\begin{eqnarray}
P_\alpha^*(\varphi) \, \sim\,  \lim\limits_{\beta = |\alpha| \to
\infty} \frac{e^{i S_\alpha(\varphi)}}{Z_{\alpha}} \, \sim\,
\delta(\varphi - \varphi^*) \, . \label{eq:statP}
\end{eqnarray}
In order to study the Langevin-time dependence of averages of an
observable $O(\varphi)$ in the limit of large $\beta = |\alpha|$,
we consider the difference with respect to the stationary
solution, i.e.\
\begin{eqnarray}
\Delta O(\varphi;\vartheta) &\equiv& \int {\rm d} \varphi\,
O(\varphi) \Delta P_\alpha(\varphi;\vartheta)\, , \label{eq:obsav}
\end{eqnarray}
where $\Delta P_\alpha \equiv P_\alpha - P_\alpha^*$ for properly
normalized distributions. Since for the classical fixed point the
limiting distribution (\ref{eq:statP}) has a compact support away
from the boundaries of integration, and assuming 
analyticity\footnote{See, however, the discussion in 
Ref.~\cite{Damgaard:1987rr}}
in $\varphi$, we may use partial integration to
write:
\begin{eqnarray}
- \lefteqn{\int {\rm d} \varphi\, O(\varphi) H_{FP}(\varphi)
\Delta P_\alpha(\varphi;\vartheta) =} \nonumber\\ && \int {\rm d}
\varphi\, \left( \frac{\partial^2 O(\varphi)}{\partial \varphi^2}
+ i\, \frac{\partial O(\varphi)}{\partial \varphi}\frac{\partial
S_\alpha(\varphi)}{\partial \varphi} \right) \Delta
P_\alpha(\varphi;\vartheta) \, . \label{eq:parint}
\end{eqnarray}
Since we are interested in plaquette averages, we consider again
$O(\varphi) = e^{i l \varphi}$ with integer $l$ and
$S_\alpha(\varphi)$ given by Eq.~(\ref{eq:acaplha}). According to
the Fokker-Planck equation (\ref{eq:contfp}) the Langevin-time
evolution for the observable average (\ref{eq:obsav}) is then
described by
\begin{eqnarray}
\frac{\partial}{\partial \vartheta}\, \Delta O(\varphi;\vartheta)
&=& - \left( l^2 + l \alpha \right) \Delta O(\varphi;\vartheta) +
l \beta \int {\rm d} \varphi\, \sin(\varphi) O(\varphi) \Delta
P_\alpha(\varphi;\vartheta)
\nonumber\\
&\simeq& - \left( l^2 + l \left[\alpha - \beta
\sin(\overline{\varphi}) \right]\right) \Delta
O(\varphi;\vartheta) \, . \label{eq:oev}
\end{eqnarray}
For the last approximate relation we used that with
Eq.~(\ref{eq:statP}) the integrand is peaked near the classical
fixed point such that an appropriate constant $\overline{\varphi}$
may be found with $\overline{\varphi} \simeq \varphi^*$ in order
to simplify the remaining integral and to get a closed equation
for $\Delta O(\varphi;\vartheta)$. Eq.~(\ref{eq:oev}) has to be
evaluated for $\beta = |\alpha|$ but in the notation we keep
$\beta$ and $\alpha$ separately for further discussion. The
Langevin-time dependence of the observable average is then given
by
\begin{eqnarray}
\Delta O(\varphi;\vartheta) &\sim& e^{- \left( l^2 + l
\left[\alpha - \beta \sin(\overline{\varphi})\right] \right)\,
\vartheta} \, . \label{eq:sol}
\end{eqnarray}
With
\begin{eqnarray}
\sin(\overline{\varphi}) = \sin(\overline{\varphi}_R)
\cosh(\overline{\varphi}_I) + i \cos(\overline{\varphi}_R)
\sinh(\overline{\varphi}_I)
\end{eqnarray}
one obtains from Eq.~(\ref{eq:sol}) a convergent result if
\begin{eqnarray}
l^2 + l \left[\alpha - \beta \sin(\overline{\varphi}_R)
\cosh(\overline{\varphi}_I) \right]  > 0 \, . \label{eq:condconv}
\end{eqnarray}
For $\beta = |\alpha| \to \infty$ we have $\overline{\varphi} =
\varphi^*$ at sufficiently large $\vartheta$. At the classical
fixed point $\partial S/\partial \varphi = \alpha - \sin
(\varphi^*) = 0$ which leads to $\lim_{\vartheta \to \infty}
\Delta O(\varphi;\vartheta) \sim \lim_{\vartheta \to \infty} e^{-
l^2 \, \vartheta} = 0$. Accordingly, the stochastic method is
expected to converge well to the stationary solution in this case,
which we indeed observe from the full solution of the Langevin
equation as described above in Sections~\ref{sec:stochre} and
\ref{sec:fp}.

We note that the solution (\ref{eq:sol}) happens to reflect
important qualitative properties of the above discussed results
also for finite $\beta$ and for $\beta \not = |\alpha|$. The
partial integration leading to Eq.~(\ref{eq:parint}) is based on
analyticity arguments and a compact distribution for the
stochastic variable away from the boundaries of integration. The
latter is also assumed for the step to the second line of
Eq.~(\ref{eq:oev}). Of course, in case the dynamics is governed by
a classical fixed point, i.e.\ if the distribution of the
dynamical variable is centered around $\overline{\varphi}$ with
$\partial S/\partial \varphi (\varphi=\overline{\varphi}) \simeq
0$, the condition (\ref{eq:condconv}) is automatically fulfilled
for any $\beta$ or $\alpha$ because of the fixed point condition
(\ref{eq:fpc}). However, fluctuations often play an important role
and to analytically argue that the Langevin flow not only
converges but converges to the correct value is more involved if
the dynamics is not governed by a classical fixed point.

The importance of fluctuations can be observed, e.g., from
Fig.~\ref{fig:expiphi}, where for $\beta \lesssim {\cal O}(10)$
substantial deviations from the classical fixed point value occur.
If $\partial S/\partial \varphi (\varphi=\overline{\varphi}) \not
= 0$ then the condition (\ref{eq:condconv}) can be formally
written as
\begin{eqnarray}
\sin(\overline{\varphi}_R) \cosh(\overline{\varphi}_I) < \frac{l +
\alpha}{\beta}  \label{eq:cond2}
\end{eqnarray}
for $l \not = 0$. This suggests that convergent results might be
difficult to obtain for $\beta \gg \alpha$, which is indeed in
accordance with our findings of Sec.~\ref{sec:stochre}. For
instance, in Fig.~\ref{fig:betadep} one observes accurate results
for $l = 1$ from the stochastic method if $\beta \lesssim \alpha =
1$, but a failure of the method for larger $\beta$. This coincides
with the fact that for $\beta \lesssim \alpha$ we find a rather
compact distribution for $\varphi$ as exemplified in
Fig.~\ref{fig:drift2}. In this case one observes that the
condition (\ref{eq:cond2}) is approximately verified if
$\overline{\varphi}_R$ and $\overline{\varphi}_I$ are allowed to
take on all $\varphi$-values of significant support. In contrast,
for $\beta$ larger than $\alpha$ we find rather wide distributions
as exemplified in Fig.~\ref{fig:drift1} and the assumptions
leading to (\ref{eq:cond2}) may not be justified. In these cases
it turns out that the observed distributions are also difficult to
reconcile with condition (\ref{eq:cond2}). In particular, the case
$\alpha = 0$ with $\beta = 1$ (left graph of
Fig.~\ref{fig:drift1}) for $l= -1$, which was employed in
Sec.~\ref{sec:complex} to demonstrate the failure of the
stochastic method, is clearly violating that condition. We also
verified some further details, e.g., that for $\alpha = 1$ and
$\alpha = 2$ the observable averages with $l=\pm 1$ and $l = \pm
2$ converge faster up to a prescribed statistical error if $l$ is
positive, as suggested by condition (\ref{eq:cond2}).

\subsection{ $SU(2)$ one-plaquette model}
\label{sec:su2oneplaquette}

\subsubsection{Direct integration}

We consider a theory where the action with real coupling parameter
$\beta$,
\begin{eqnarray}
S(U) = \frac{\beta}{2}\, {\rm Tr}\, U \, ,\label{eq:ac2onepl}
\end{eqnarray}
is invariant under the symmetry transformation
\begin{eqnarray}
U \rightarrow W^{-1} U W \label{eq:su2sym}
\end{eqnarray}
with $U,W \in SU(2)$. For the analytical calculations we use a
parametrization of $SU(2)$ matrices in terms of an angle $\varphi$
and a unit vector $\vec{n}$:
\begin{eqnarray}
U\left(\varphi,\vec{n}\right) \, =\, e^{i \varphi \vec{n} \cdot
\vec{\sigma}/2} \, =\, \left( \cos { \varphi \over 2 } \right) {
\bf 1 } + i \left( \sin { \varphi \over 2 } \right) \vec{n} \cdot
\vec{\sigma} \, , \label{eq:parametsu2}
\end{eqnarray}
where $ 0 \le \varphi < 2\pi$ with the three Pauli matrices
$\vec{\sigma}$. Using the Haar measure in terms of these variables
we obtain averages of an observable $O(U)$ by direct integration
from
\begin{eqnarray}
\langle O(U) \rangle &=& \frac{1}{Z} \int {\rm d} U\, e^{i S(U)}\, O(U) \nonumber\\
&=& \frac{1}{Z} \int_0^{2 \pi} {\rm d} \varphi \int \frac{{\rm d}
\Omega(\vec{n})}{4\pi}\, \left( \sin\frac{\varphi}{2}\right)^2
\e^{i \beta \cos(\varphi/2)}\,
O\left(U\left(\varphi,\vec{n}\right)\right) \, ,
\end{eqnarray}
where $\Omega(\vec{n})$ is the uniform measure on the unit sphere
and
\begin{eqnarray}
Z &=& \int_{0}^{2\pi}{\rm d}\varphi \, \left(
\sin\frac{\varphi}{2}\right)^2 \e^{i \beta \cos(\varphi/2)} \, =
\, \frac{2 \pi}{\beta}\, J_1(\beta)  \, . \label{eq:rhm}
\end{eqnarray}
For instance, with ${\rm Tr}\, U/2 = \cos(\varphi/2)$ the
plaquette average as a function of $\beta$ is
\begin{eqnarray}
\left\langle {1\over 2 }{\rm Tr}\, U \right\rangle &=& i\,
\frac{J_2(\beta)}{J_1(\beta)} \, . \label{eq:exsu2}
\end{eqnarray}
Note that for this particular observable the value of the integral
coincides with the one of the plaquette average in
Eq.~(\ref{eq:avu1}) for the ($\alpha = 1$)-reweighted $U(1)$
model, and the analytic results are plotted as solid curves in
Fig.~\ref{fig:betadep}.

\subsubsection{Complex Langevin equation}

We will compare results obtained from direct integration to
estimates from a stochastic process. As described in
Sec.~\ref{sec:realtime}, this requires an extension of the
original $SU(2)$ manifold to $SL(2,{\bf C})$ for the Langevin
dynamics. Writing
\begin{eqnarray}
U \,=\, a\, {\bf 1} + i \vec{b}\cdot \vec{\sigma} \,=\,
\left(\begin{array}{cc} a+ i b_3 & b_2+ i b_1 \\ -b_2 + i b_1 &a -
i b_3 \end{array} \right) \label{eq:param2}
\end{eqnarray}
with ${\rm det}\,U = a^2+\vec{b}^2=1$ the coefficients $a$ and
$\vec{b}$ are complex numbers for $U \in SL(2,{\bf C})$. The
Langevin equation follows from Eq.~(\ref{eq:ALangevinM}). We
expand the exponential in that equation to first order in
$\epsilon$, which means we must include the square of the noise
term, which is proportional to unity. The evolution equation then
reads for the one-plaquette model with action (\ref{eq:ac2onepl}):
\begin{eqnarray}
U^{\prime} &=& \left( \tilde{a}\, {\bf 1}   + i \sum_{a=1}^{3}
\sigma_a \left(- \epsilon\, \frac{\beta}{2}\,{\rm Tr}(\sigma_a U)
+ \sqrt{\epsilon}\, \eta_a \right) \right) U\, ,
\label{eq:langsu2}
\end{eqnarray}
with Gaussian noise $\eta_{a}$ corresponding to
Eq.~(\ref{eq:realtimenoise}). In order to stay in group space the
constant $\tilde{a}$ in this equation is calculated from
\mbox{$\tilde{a} = \sqrt{1 - (- \epsilon\beta{\rm Tr}(\sigma_a
U)/2 + \sqrt{\epsilon}\, \eta_a)^2}$}. Alternatively, one can also
appropriately normalize that matrix. In contrast to the procedure
of Sec.~\ref{sec:u1}, Eq.~(\ref{eq:langsu2}) describes the
stochastic dynamics directly in terms of group elements $U$. This
is closer to what we will do for the $SU(2)$ gauge theory in $3+1$
dimensions in Sec.~\ref{sec:su2gaugetheory}, and is used to
introduce some concepts that will be employed for the field theory
as well.

\begin{figure}[t]
\begin{center}
\epsfig{file=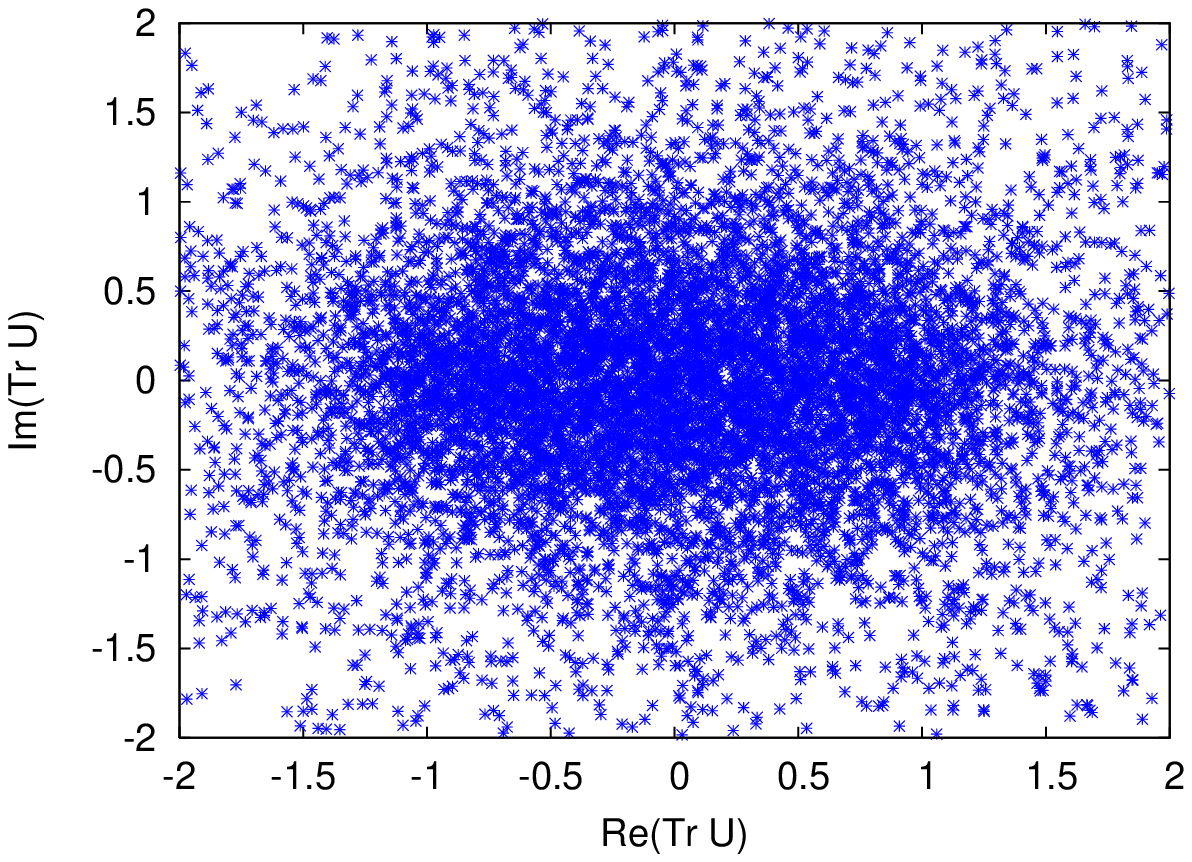,width=6.7cm}
\epsfig{file=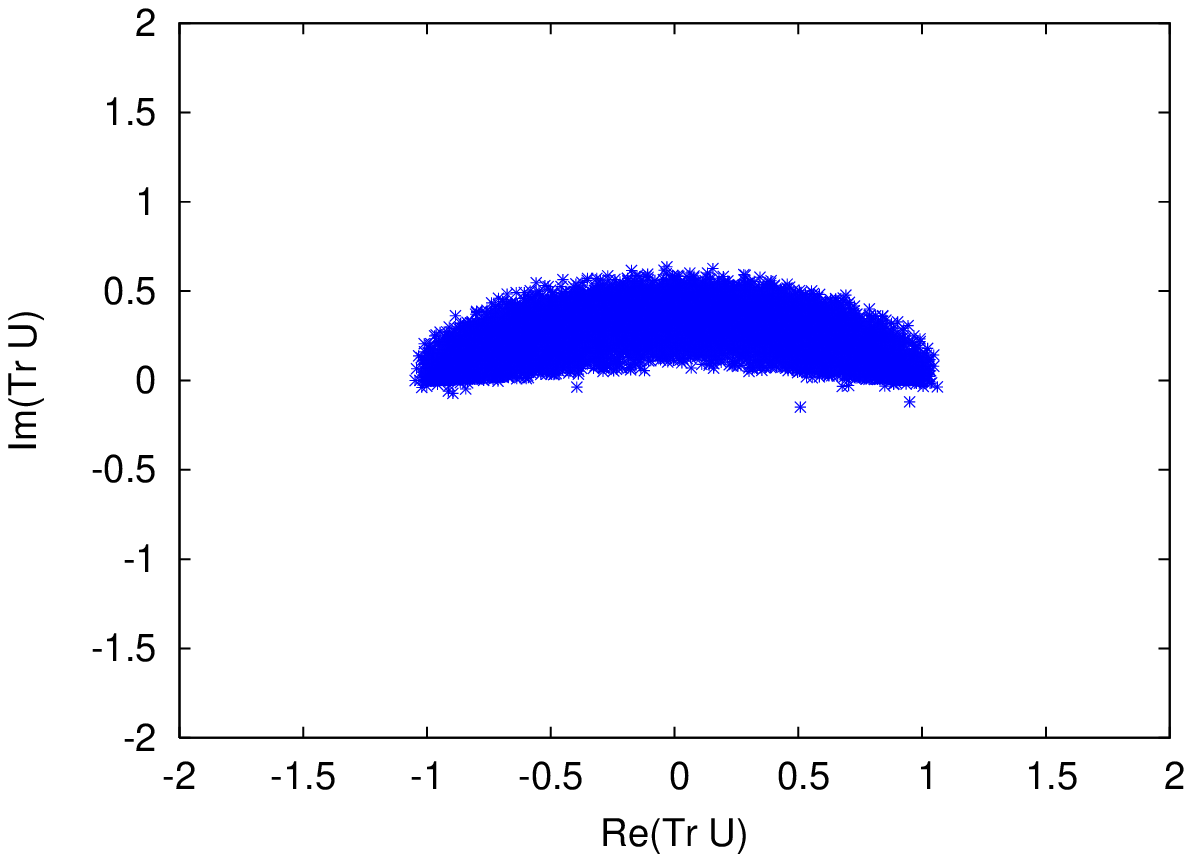,width=6.7cm}
\end{center}
\vspace*{-0.5cm} \caption{\small The real and imaginary part of
the plaquette variable ${\rm Tr} U$ from snapshots with constant
Langevin-time stepping for the $SU(2)$ one-plaquette model with
$\beta = 1$. The left graph shows the wide distribution of values
obtained from the standard Langevin dynamics, while the right
graph displays the compact distribution from the ("gauge-fixed")
optimized process (see text for explanation).} \label{fig:distsu2}
\end{figure}
Applying the stochastic process to a computation of the plaquette
average for $\beta =1$ yields a result consistent with zero:
\begin{eqnarray}
\left\langle \frac{1}{2}\, {\rm Tr}\, U \right\rangle &
\stackrel{\rm without \atop optimization}{=}& -0.02(\pm 0.02) -
i\, 0.01(\pm 0.02) \, , \label{eq:wsu2}
\end{eqnarray}
which disagrees with the non-vanishing exact value $i
J_2(1)/J_1(1)\simeq i\, 0.261$ given by Eq.~(\ref{eq:exsu2}). The
situation is analogous to the one described in Sec.~\ref{sec:u1}.
In particular, the distribution for the stochastic variable
obtained from the solution of the Langevin equation
(\ref{eq:langsu2}) exhibits similar qualitative patterns. For
instance, the left graph of Fig.~\ref{fig:distsu2} shows for
$\beta = 1$ the distribution of ${\rm Tr}\, U$ in the complex
plane. One observes a wide distribution, reminiscent of the left
graph shown in Fig.~\ref{fig:drift1} for the $U(1)$ model without
optimized updating.

\subsubsection{Optimized updating by "gauge fixing"}
\label{subsec:optimized}

An optimized updating scheme for the Langevin dynamics with the
aim to calculate accurate results may be achieved in various ways.
Motivated by the results of Sec.~\ref{sec:u1}, the optimized
updating should control the growth of fluctuations for the complex
Langevin equation. Since the $SU(2)$ one-plaquette model has a
global symmetry described in Eq.~(\ref{eq:su2sym}), which is
reminiscent of a local gauge transformation in the corresponding
field theory, one may use this symmetry to "gauge-fix" certain
variables in order to constrain the growth of fluctuations. For
the complex Langevin equation the plaquette variable $U$ as well
as the matrix $W$ in Eq.~(\ref{eq:su2sym}) are elements of
$SL(2,{\bf C})$. In the following we will use it in order to
diagonalize $U$ after each successive Langevin-time step. For the
representation (\ref{eq:parametsu2}) this corresponds to the
``gauge-condition'' $\vec{n} = (0,0,1)$, or, with the
parameterization of Eq.~(\ref{eq:param2}) we write after each
Langevin updating step $U^{\prime} \,=\, {\rm diag}\left(a+ i
\sqrt{1-a^2}, a - i \sqrt{1-a^2}\right)$, which corresponds to
choosing $\vec{b} = ( 0, 0,
\sqrt{1-a^2})$.\footnote{Alternatively, one can choose the
negative sign with $\vec{b} = ( 0, 0, -\sqrt{1-a^2})$, which makes
no difference for the following discussion.}

\begin{table}
\begin{center}
\begin{tabular}{|c||c|c||c|c|}
\hline
observable & exact Re &   stochastic Re & exact Im & stochastic Im  \\
\hline\hline
$ \langle {\rm Tr} U/2 \rangle$ & 0 & $-0.004 (\pm 0.006)$ & $0.261$ & $ 0.260 (\pm 0.001) $ \\
\hline $ \langle ( {\rm Tr} U/2 )^2 \rangle $ &  $0.216$ &  $
0.217 (\pm 0.003) $ & 0 &
  $ -0.001 (\pm  0.002) $  \\
\hline
\end{tabular}
\end{center}
\caption{\small Results from the optimized stochastic process, as
described in Sec.~\ref{subsec:optimized}, are compared to exact
values from direct integration for $\beta = 1$. For the latter
only three significant digits are given. Listed are results for
the real and imaginary part of $\langle \Tr U/2 \rangle$ and
$\langle (\Tr U/2)^2 \rangle$ denoted as Re and Im, respectively.}
\label{tab:tabsu2}
\end{table}
Using this procedure we find that, in contrast to the wrong
results displayed in Eq.~(\ref{eq:wsu2}) for $\beta = 1$, the
optimized stochastic process now reproduces correct averages. This
is exemplified in Table~\ref{tab:tabsu2} for two different
observables, $\langle {\rm Tr} U/2 \rangle$ and $\langle ({\rm Tr}
U/2)^2 \rangle$. One can compare the distribution of values
obtained from snapshots at equidistant Langevin-time steps with
and without optimized updating. The right graph of
Fig.~\ref{fig:distsu2} shows that the "gauge-fixing" leads to a
compact distribution, in contrast to the one from the unmodified
process displayed on the left of that figure. Moreover, for
$\langle {\rm Tr} U/2 \rangle$ as a function of $\beta$ we obtain
a very similar plot than the one shown in Fig.~\ref{fig:betadep}
with accurate results for $\beta \lesssim 1$ and wrong results for
somewhat larger $\beta$. Of course, according to
Eqs.~(\ref{eq:avu1}) and (\ref{eq:exsu2}), the expectation values
have to coincide in this particular case in the range where
correct values are obtained.

We note that the discussion can be performed along very similar
lines than what has been done in Sec.~\ref{sec:u1} for the $U(1)$
one-plaquette model. This can be observed from the fact that the
above optimized updating procedure can be mapped to a Langevin
process with one degree of freedom only, given by
\begin{eqnarray}
\varphi^\prime = \varphi - i \epsilon\, \frac{\beta}{2}
\sin\frac{\varphi}{2} + \epsilon\, \cot\frac{\varphi}{2} +
\sqrt{\epsilon}\, \eta \, \label{eq:actionmodsu2}
\end{eqnarray}
with white noise $\eta$. This Langevin equation follows directly
from Eq.~(\ref{eq:rhm}) by writing the reduced Haar measure as an
exponential. The equivalence with the optimized updating is
verified explicitly in an expansion of Eq.~(\ref{eq:langsu2}) to
order $\epsilon$ in the appendix.

In contrast to the $(\alpha=\beta)$-reweighted one-plaquette model
of Sec.~\ref{sec:u1}, the $SU(2)$ one-plaquette model has no
well-defined limit $\beta \to \infty$, in which fluctuations are
suppressed. This can be observed, for instance, from the integral
in Eq.~(\ref{eq:exsu2}). For the following discussion it will be
an important property that the $SU(2)$ gauge theory has a
well-defined limit $\beta \to \infty$, which corresponds to the
continuum limit of the lattice theory.

\section{$SU(2)$ gauge theory in $3+1$ dimensions}
\label{sec:su2gaugetheory}

\subsection{Stochastic dynamics without optimized updating}

Real-time simulations for quantum field theories at non-zero
temperature require a complex time-path, where the imaginary-time
extent is given by the inverse temperature $\sim 1/T$ and the
physical time determines the real-time extent. For scalar and
$SU(2)$ gauge theory this has been investigated in
Ref.~\cite{Berges:2006xc} to which we refer for further details.
Here we re-consider $SU(2)$ gauge theory. In this case the
real-time quantum dynamics in 3+1 dimensions is obtained from a
stochastic process in the additional (5th) Langevin-time according
to Eq.~(\ref{eq:ALangevinM}) explained in Sec.~\ref{sec:realtime}.
The numerical results will be obtained using equal couplings for
the time-like and space-like plaquettes in Eq.~(\ref{eq:ganisoM}),
i.e.\ $g_0^2 = g_s^2 \equiv g^2 \sim 1/\beta$ with $N_s=4$ and $N_t=8$. 
All values are
given in units of appropriate powers of the spatial lattice
distance $a_s$.

\begin{figure}[t]
\begin{center}
\epsfig{file=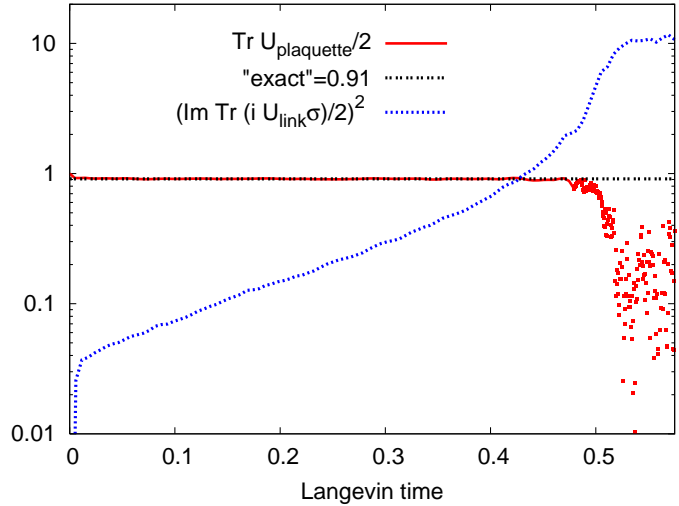,width=10.cm}
\end{center}
\vspace*{-0.5cm} \caption{\small The (red) solid line shows the
plaquette variable ${\rm Tr}\, U_{\rm plaquette}/2$ defined in
Eq.~(\ref{eq:plaquettevariable}) as a function of Langevin-time
using $g=0.5$. We employ a complex contour with non-zero real-time
extent at temperature $T=1$ as explained in the text. For
comparison, the (black) dashed line gives the corresponding
"exact" result for a Euclidean field theory at the same
temperature. Since ${\rm Tr}\, U_{\rm plaquette}/2$ is
time-independent, the Langevin-time averages of both results have
to agree. The deviation at late Langevin-times is signalled by
large values for the squared link variable $\left({\rm Im}\, {\rm
Tr}\, (i U_{\rm link}\, \vec{\sigma})/2\right)^2$ defined in
Eq.~(\ref{eq:ulink}), which is displayed as a (blue) dotted
curve.} \label{fig:su2nogaugefixing}
\end{figure}
In Ref.~\cite{Berges:2006xc} it was shown that without
optimization the Langevin dynamics described by
Eq.~(\ref{eq:ALangevinM}) fails to reproduce correct results in
the limit of large Langevin-time. It particular, it was seen that
the Langevin flow approaches the correct results at intermediate
Langevin-times before it finally starts deviating. This is
illustrated in Fig.~\ref{fig:su2nogaugefixing} for the gauge
invariant spatial plaquette averaged on the lattice:
\begin{eqnarray}
\frac{1}{2}\, {\rm Tr}\, U_{\rm plaquette} \equiv \frac{1}{6 N_s^3
N_t} \sum_{x,\, i<j}  {\rm Tr}\, U_{x,ij} \, ,
\label{eq:plaquettevariable}
\end{eqnarray}
where the plaquette variable $U_{x,\mu \nu}$ is defined in
Eq.~(\ref{eq:plaq}). The (red) solid line shows the result as a
function of the Langevin-time for a complex (isoceles) triangle
contour with real-time extent $\Delta t_R = 1$ and imginary-time
extent $\Delta t_I = 1$, corresponding to a thermal theory at
temperature $T=1$.\footnote{See Ref.~\cite{Berges:2006xc} for a
discussion of complex time contours in this context.} For
comparison, the (black) dashed line gives the "exact" value for
this observable as obtained from stochastic quantization in {\em
Euclidean} space time, i.e.\ for a time-contour with no real-time
extent. In the latter case one can prove the convergence of the
stochastic method~\cite{Damgaard:1987rr}, and for the employed
parameters we find at late Langevin-times ${\rm Tr}\, U_{\rm
plaquette}/2 = 0.91$ giving two significant digits. Of course,
this comparison is only possible because we consider the special
case of a time-independent observable, whose value has to be the
same in Euclidean as well as Minkowskian space-time. From
Fig.~\ref{fig:su2nogaugefixing} we see that the result is indeed
independent of the employed time-path for not too late
Langevin-times. However, finally deviations occur demonstrating
the breakdown of the complex Langevin method. In this case the
plaquette variable (\ref{eq:plaquettevariable}) develops 
larger fluctuations of the imaginary part, whose Langevin-time average
is zero, and in Fig.~\ref{fig:su2nogaugefixing} we only show
its real part. The onset time for deviations can be delayed by
further decreasing the real-time extent of the lattice, which is
analyzed in detail in Ref.~\cite{Berges:2006xc}. In principle this
may be used to extract physical results for sufficiently short
real times, however, this procedure would provide severe
restrictions for actual applications of the method.

A characteristic measure that may be used to monitor this
breakdown is given by the quantity
\begin{equation}
\frac{1}{4} \left({\rm Im}\, {\rm Tr}\, (i U_{\rm link}\,
\vec{\sigma})\right)^2 \equiv \frac{1}{4 N_s^3 N_t} \sum_{x,j}
\left({\rm Im}\, {\rm Tr}\, (i U_{x,j}\,\vec{\sigma} )\right)^2 =
\frac{1}{ N_s^3 N_t} \sum_{x,j} \left({\rm Im}\, \vec{b}_{x,j}
\right)^2 , \label{eq:ulink}
\end{equation}
where we used for the last equation the representation
corresponding to Eq.~(\ref{eq:param2}). This quantity is not gauge
invariant and would vanish identically for $U \in SU(2)$. As
explained in Sec.~\ref{sec:realtime}, for the complex Langevin
dynamics $U \in SL(2,{\bf C})$ and, therefore, the quantity
$(\ref{eq:ulink})$ can be non-zero and provides a characteristic
quantity to measure deviations from $SU(2)$. From the (blue)
dotted line of Fig.~\ref{fig:su2nogaugefixing} one observes that
the breakdown of the complex Langevin dynamics occurs once this
quantity becomes significantly larger than one.

\begin{figure}[t]
\begin{center}
\epsfig{file=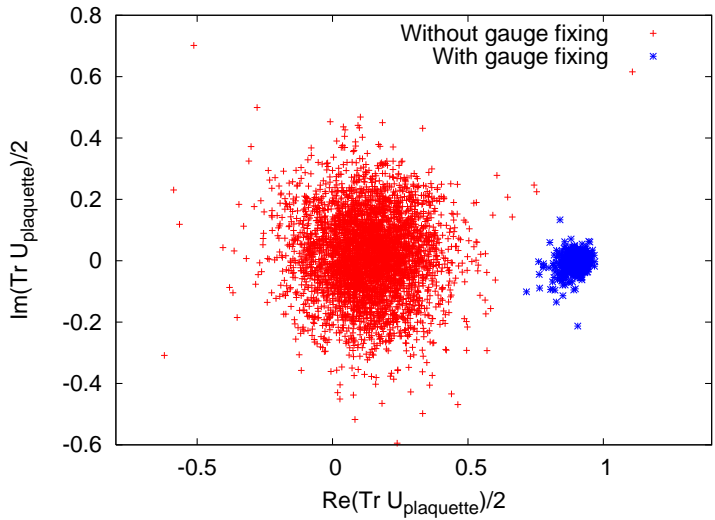,width=10.cm}
\end{center}
\vspace*{-0.5cm} \caption{\small Distributions of the values for
the averaged spatial plaquettes ${\rm Tr}\, U_{\rm plaquette}/2$
in the complex plane from snapshots of the Langevin process. The
(red) crosses give the distribution for large Langevin-times
obtained from the stochastic process without optimization. The
(blue) stars give the distribution for large Langevin-times from
optimized updating using gauge fixing, which yields correct
observable averages. The employed parameters correspond to those
used in Figs.~\ref{fig:su2nogaugefixing} and \ref{fig:smallg}. }
\label{fig:distributionsu2}
\end{figure}
In order to make contact with the discussions of
Secs.~\ref{sec:u1} and \ref{sec:su2oneplaquette}, we display in
Fig.~\ref{fig:distributionsu2} the real and imaginary part of the
plaquette variable (\ref{eq:plaquettevariable}) from snapshots
with constant Langevin-time stepping. The employed parameters are
the same as for Fig.~\ref{fig:su2nogaugefixing}. The (red) crosses
give the distribution for sufficiently large Langevin-times, i.e.\
for times when the plaquette variable deviates from the correct
results. From Fig.~\ref{fig:su2nogaugefixing} one observes that
for the employed parameters this is the case for $\vartheta
\gtrsim 0.5$. For these times Fig.~\ref{fig:distributionsu2}
exhibits a relatively widespread distribution of values in the
complex plane. This is similar to what is observed in
Secs.~\ref{sec:u1} and \ref{sec:su2oneplaquette} for the $U(1)$
and $SU(2)$ one-plaquette models for those cases where the complex
Langevin method fails. In contrast, for earlier Langevin-times we
find a compact distribution similar to what is displayed as (blue)
stars in Fig.~\ref{fig:distributionsu2}. Accordingly, for these
earlier Langevin-times the values for the plaquette variable
(\ref{eq:plaquettevariable}) agree well with accurate results. The
crucial role of a compact distribution for the convergence to
accurate results was also observed for the one-plaquette models in
Secs.~\ref{sec:u1} and \ref{sec:su2oneplaquette}. In the
following, we will show how to stabilize a compact distribution
for all Langevin-times using optimized updating by gauge fixing.
The (blue) stars in Fig.~\ref{fig:distributionsu2} actually
correspond to results obtained from gauge-fixed Langevin dynamics,
which will be explained below.

\subsection{Optimized updating by gauge fixing}

In the previous sections we have observed that the breakdown of
the complex Langevin dynamics occurs in the presence of large
fluctuations of the complexified dynamical variables. In the
spirit of Sec.~\ref{sec:su2oneplaquette}, we consider here a
reduction of fluctuations by gauge fixing. We employ maximal axial
gauge, i.e.\ on a periodic lattice\footnote{The time contour is
also periodic because we are studying thermal equilibrium.} one
can fix by gauge transformations
\begin{eqnarray}
U_{x,\mu} \rightarrow W^{-1}(x) U_{x,\mu} W(x + \hat{\mu})
\label{eq:su2gauge}
\end{eqnarray}
the link variables to one for the following links:
\begin{eqnarray}
  &&  \mu=0,\,\,  0 \le x^0 < N_t-1,\,\, 0 \le x^1,x^2,x^3 < N_s\, ,  \nonumber\\
  &&  \mu=1,\,\,  0 \le x^1 < N_s-1,\,\, 0 \le x^2,x^3 < N_s,\,\, x^0=0\, ,  \nonumber\\
  &&  \mu=2,\,\,  0 \le x^2 < N_s-1,\,\, 0 \le x^3 < N_s,\,\, x^0=x^1=0\, , \nonumber\\
  &&  \mu=3,\,\,  0 \le x^3 < N_s-1,\,\, x^0=x^1=x^2=0 \, ,
  \label{eq:fixing}
\end{eqnarray}
where the lattice points are labelled by integers $0\le
x^1,x^2,x^3 < N_s$ and $0 \le x^0 < N_t$. The gauge-fixed links
are not updated~\cite{montvay} and the rest is updated according
to Eq.~(\ref{eq:ALangevinM}).

Another possibility, which would be the analogue of what was
employed for the one-plaquette model in
Sec.~\ref{sec:su2oneplaquette}, is to update all links but after
each Langevin-time step to calculate and apply the field of gauge
transformations $W(x)$ in order to fix the link variables
according to Eq.~(\ref{eq:fixing}). One can build up the field of
gauge transformations by fixing first $W(0,0,0,0)={\bf 1}$, than
get the values $W(0,0,0,z)$ by solving $W^{-1}(x) U_{x,\mu} W(x +
\hat{\mu}) = {\bf 1}$ according to the last line of
Eq.~(\ref{eq:fixing}), than the values $W(0,0,y,z)$ according to
the third line, and so on. The latter method turns out to be not
as efficient in suppressing fluctuations as the maximal axial
gauge.

\begin{figure}[t]
\begin{center}
\epsfig{file=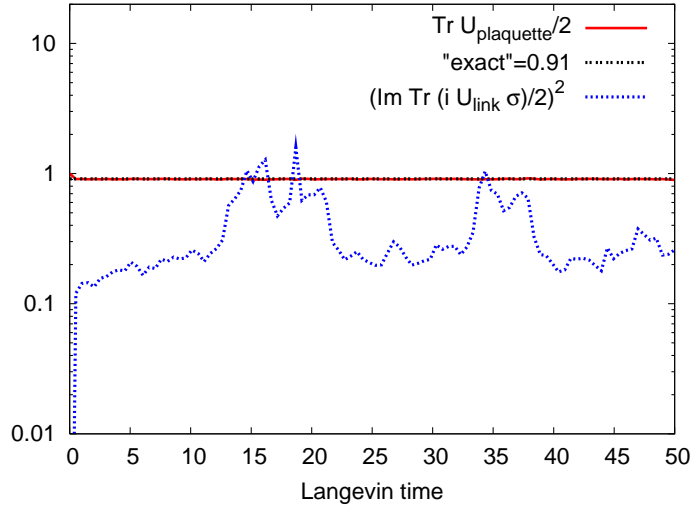,width=10.cm}
\end{center}
\vspace*{-0.5cm} \caption{\small Same as
Fig.~\ref{fig:su2nogaugefixing} but with gauge fixing. In contrast
to the former, the results from the complex Langevin equation
agree well to the "exact" values even at large Langevin-times.
Note that the displayed Langevin-time exceeds the one of
Fig.~\ref{fig:su2nogaugefixing} by about a factor of $100$.}
\label{fig:smallg}
\end{figure}
In Fig.~\ref{fig:smallg} we show the results corresponding to
Fig.~\ref{fig:su2nogaugefixing}, however, now obtained from the
stochastic dynamics with optimized updating by gauge fixing. One
observes that the result from the complex Langevin equation stays
close to the "exact" result and we found no sign of increasing
deviations at sufficiently large Langevin-times. Accordingly, also
the quantity (\ref{eq:ulink}) stays comparably small as displayed
in Fig.~\ref{fig:smallg}. This reflects the fact that the
distribution of the dynamical variable in the complex plane
remains relatively compact, which is exemplified in
Fig.~\ref{fig:distributionsu2} for the plaquette variable
(\ref{eq:plaquettevariable}).

This is an important advance as compared to the results of
Ref.~\cite{Berges:2006xc}, where no stable physical fixed point
solution of the Langevin equation could be observed. Gauge fixing
turns out to be an efficient way to reduce fluctuations of the
complex Langevin dynamics. However, increasing the coupling leads
to increased fluctuations and we find that gauge fixing alone is
not enough for sufficiently large $g$. Therefore, the combination
of gauge fixing and not too large values for the coupling lead to
the quantitative results we observe. For the results of
Fig.~\ref{fig:smallg} we use $g = 0.5$. For the employed
parameters with temperature $T=1$, we find that increasing $g$ to
values larger than about one leads to deviations from correct
results at large Langevin-times similar to the situation displayed
in Fig.~\ref{fig:su2nogaugefixing}. 
Decreasing the temperature or using shorter real-time extent improves 
the situation, as has been found also without gauge fixing in 
Ref.~\cite{Berges:2006xc}.
We have checked on $N^3_S=4^3 , ..., 32^3$ lattices 
that the value of $g$ one has to use for physical results does not 
depend on the spatial size of the lattice. In principle, this means that
even though the small $g$ (and thus larger $\beta$) corresponds to
smaller lattice spacings, with a bigger lattice size one could overcome this 
effect at the expense of computational time. However, in practice 
this is difficult to achieve for $g<1$ because of limited resources.

The discussion is reminiscent of the one in Sec.~\ref{sec:u1},
where it is shown that the ($\alpha = \beta$) reweighted $U(1)$
one-plaquette model is governed by classical dynamics in the limit
$\beta \to \infty$.\footnote{Note that the $SU(2)$ one-plaquette
model has no well defined limit $\beta \to \infty$ as discussed in
Sec.~\ref{sec:su2oneplaquette}.} The reduction of fluctuations by
increasing $\beta$ in that model is exemplified in
Figs.~\ref{fig:width} and \ref{fig:drift2}. Qualitatively similar,
for the gauge theory one obtains for larger $\beta \sim 1/g^2$
more compact distributions than shown in
Fig.~\ref{fig:distributionsu2}.

\section{Conclusions}
\label{sec:conclusions}

For the $SU(2)$ gauge theory with action (\ref{eq:clgaugeaction})
real-time stochastic quantization requires the dynamical variables
to become elements of $SL(2,{\bf C})$. The expectation values of
the underlying gauge theory are recovered after taking noise or
Langevin-time averages, respectively. This change from the compact
$SU(2)$ gauge group to the non-compact $SL(2,{\bf C})$ has
important consequences for the Langevin dynamics. For the former
${\rm det} U = a^2+\vec{b}^2 = 1$ yields finite values for the
real $a$ and $\vec{b}$, using the representation of the link
variables corresponding to Eq.~(\ref{eq:param2}). In contrast, the
$SL(2,{\bf C})$ group admits unbounded values for the now complex
$a$ and $\vec{b}$. Accordingly, we observe that without optimized
updating the complex Langevin equation yields large fluctuations
for the dynamical variables at sufficiently late Langevin-times,
which finally lead to a breakdown of the method. Apart from the
$SU(2)$ gauge theory in $3+1$ dimensions, we observe analogous
findings for the $U(1)$ and $SU(2)$ one-plaquette models. The
simplicity of the one-plaquette models allow an analytical
analysis as presented in Secs.~\ref{sec:u1} and
\ref{sec:su2oneplaquette}.

In this paper we showed that large fluctuations of the
complexified dynamical variables can be efficiently reduced by
employing optimized updating procedures for the Langevin process.
Here we investigated optimized updating using stochastic
reweighting or gauge fixing, respectively. These procedures do not
affect the underlying theory but can stabilize the physical fixed
point of the Langevin equation. The success of stochastic
reweighting was found to be linked to the appearance of attractive
fixed points of the Langevin flow, while the gauge fixing simply
constrains the dynamical variables. For the gauge theory we
demonstrated that gauge fixing leads to an efficient reduction of
fluctuations for not too small $\beta \sim 1/g^2$: We employed
maximal axial gauge and calculated plaquette averages on a lattice
with non-zero real-time extent. Where applicable, the results were
shown to accurately reproduce alternative calculations in
Euclidean space-time. This is an important advance as compared to
the results of Ref.~\cite{Berges:2006xc}, where no stable physical
fixed point solution of the Langevin equation could be observed
for the non-Abelian gauge theory even for small couplings.

For the reweighted $U(1)$ and the "gauge-fixed" $SU(2)$
one-plaquette models, we obtained accurate results also for small
$\beta$. The fact that real-time stochastic quantization is
simpler to apply to non-gauge theories is in accordance with
earlier findings for scalar field theories in
Ref.~\cite{Berges:2006xc}. Since large $\beta$ describe the
continuum limit of the lattice gauge theory, with the present
results at hand one has in principle a procedure to do
simulations in non-Abelian gauge theory. However, without further improvements,
it is difficult to achieve $g \lesssim 1$ because of limited
computational resources.
Also further tests using different time contours along the lines of 
Ref.~\cite{Berges:2006xc} are necessary before pressing questions of 
calculations of transport properties and nonequilibrium gauge
theory dynamics are to be addressed. \\

\noindent {\bf Acknowledgements:} We are indebted to Ion-Olimpiu
Stamatescu for collaboration in an early stage of a common work on
reweighting techniques for one-plaquette models and proposing a
modified Langevin equation (\ref{eq:clangedis}). We also thank
Szabolcs Bors{\'a}nyi for helpful discussions and fruitful
collaboration on related work. This work was supported in part by the 
BMBF grant 06DA267, and by the DFG under contract SFB634.

\section{Appendix}

In this appendix we show that the optimized updating procedure
employed for the $SU(2)$ one-plaquette model in
Sec.~\ref{subsec:optimized} can be mapped to a Langevin process
with one degree of freedom only, described by
Eq.~(\ref{eq:actionmodsu2}).

Using the parameterization of Eq.~(\ref{eq:param2}) we write
$U=a+i \vec{b} \vec{\sigma}$. The "gauge fixing" condition employed in
Sec.~\ref{subsec:optimized} then reads $\vec{b} = ( 0, 0,
\sqrt{1-a^2})$. We denote in Eq.~(\ref{eq:langsu2}) the
Langevin-time-stepping matrix as $\Theta=\tilde a + i \vec{d} \vec{\sigma}
$. Because $U$ is rotated to the fixed gauge after the last
Langevin-time step we have $b_1=b_2=0$, ${\rm Tr}(\sigma_1 U)=
{\rm Tr}(\sigma_2 U)=0$ and, accordingly,
\begin{equation}\label{eq:drift_terms}
   d_1=\eta_1 \sqrt{\epsilon},\ \ \ \
   d_2=\eta_2 \sqrt{\epsilon},\ \ \ \
   d_3=-\epsilon {\beta\over 2 } {\rm Tr} (\sigma_3 U) +\eta_3
   \sqrt{\epsilon} \, .
\end{equation}
Multiplying $\Theta$ and $U$, one obtains $U'=a'+i \vec{b}' \vec{\sigma}$
using $b_1=b_2=0$ and $ (\vec{v} \vec{\sigma})(\vec{w} \vec{\sigma}) 
= \vec{v} \vec{w}+ i \vec{\sigma}  ( \vec{v} \times \vec{w})$  :
\begin{equation}
a'+i \vec{b}' \vec{\sigma}= a \tilde a - \vec{b}\vec{d} + 
           i\sigma_1 (ad_1 -d_2 b_3)
                            + i\sigma_2 (a d_2 + d_1 b_3 )
        + i\sigma_3 ( \tilde a b_3 + a d_3 ) \, .
\end{equation}
Again rotating to the fixed gauge one obtains $U''=a''+i
\vec{b}'' \vec{\sigma}$ with
\begin{equation}
 a''=a' , \ \ \ \ b_1''=b_2''=0, \ \ \ \
    b_3''= \sqrt{ b_1'^2+b_2'^2+b_3'^2 } \, .
\end{equation}
Expanding $b_3''$ to order  $ \epsilon $ according to 
Eq.~(\ref{eq:drift_terms}) leads to
\begin{equation}
 b_3'' = (1 - {d_3^2 \over 2} ) b_3 +
 a \left( d_3 + { d_1^2 + d_2^2 \over 2 } {a \over b_3} \right) \,
 .
\end{equation}
One can write $d_1^2+d_2^2= 4 \epsilon + \epsilon \rho $, where
$\rho$ is a noise term with zero mean. Since it is multiplied with
$\epsilon$, in the continuum limit one can neglect this
term.\footnote{For finite continuum contribution it would need to
scale with $\sqrt{\epsilon}$.} Using the notation $b_3''= \sin(
\varphi'' / 2 )$ and similarly $ b_3= \sin( \varphi / 2 )$,  and $
\Delta \varphi= d_3 + 2 \epsilon  {a\over b_3} $, and noticing
that to order $\epsilon$ one has $ \cos \Delta \varphi =(1 -
{d_3^2 \over 2} )$ yields
\begin{equation}
 \sin{\varphi''\over 2 } = \cos (\Delta \varphi) \sin {\varphi
\over 2}
 + \cos{\varphi\over 2} \sin (\Delta \varphi) \, .
\end{equation}
Since this describes angle addition, we get:
\begin{equation}
 {\varphi''\over 2 }-  {\varphi\over 2} = d_3 + 2 \epsilon
{a\over b_3}
 =  -i \epsilon \beta \sin {\varphi\over 2} + 2 \epsilon
\cot {\varphi \over 2} + \eta_3 \sqrt{\epsilon} \, ,
\end{equation}
which is the discretised Langevin equation (\ref{eq:actionmodsu2})
as obtained from the ``reduced'' action $S = 2 \ln \sin
{\varphi /  2} + i \beta \cos {\varphi / 2}$, with
Langevin-time step $4 \epsilon$.


\begin{thebibliography}{10}

\vspace*{0.35cm}

\bibitem{stochquant} G.~Parisi and Y.-S.~Wu , Sci.\ Sin., Ser.\ A, Math.\ Phys.\ Astron.\ Tech.\ Sci.\ {\bf 24} (1981) 483.

\bibitem{Damgaard:1987rr}
  P.~H.\ Damgaard and H.~H\"uffel,
  Phys.\ Rept.\  {\bf 152} (1987) 227.

\bibitem{Berges:2005yt}
J.~Berges and I.~O.~Stamatescu, Phys.\ Rev.\ Lett.\ {\bf 95}
(2005) 202003 [arXiv:hep-lat/0508030].

\bibitem{Berges:2006xc}
  J.~Berges, S.~Borsanyi, D.~Sexty and I.~O.~Stamatescu,
  Phys.\ Rev.\  D {\bf 75} (2007) 045007
  [arXiv:hep-lat/0609058].

\bibitem{cl}
J.~R.~Klauder, in {\em Recent developments in High Energy
Physics}, edited by H.~Mitter and C.~B.~Lang (Springer, New York,
1983); Phys.\ Rev.\ A {\bf 29} (1984) 2036. G.~Parisi, Phys.\
Lett. B {\bf 131} (1983) 393.

\bibitem{Minkowski}
H.~H\"uffel and H.~Rumpf, Phys.\ Lett.\ B {\bf 148} (1984) 104.
E.~Gozzi,
Phys.\ Lett.\ B {\bf 150} (1985) 119.
D.~J.~E.~Callaway, F.~Cooper, J.~R.~Klauder and H.~Rose,
Nucl.\ Phys.\ B {\bf 262} (1985) 19.
H.~Nakazato and Y.~Yamanaka,
Phys.\ Rev.\ D {\bf 34} (1986) 492.
H.~H\"uffel and P.~V.~Landshoff,
Nucl.\ Phys.\ B {\bf 260} (1985) 545.
K.~Okano, L.~Schulke and B.~Zheng,
Prog.\ Theor.\ Phys.\ Suppl.\  {\bf 111} (1993) 313.

\bibitem{convergence}
J.~Klauder and W.~Petersen, J.\ Stat.\ Phys.\ {\bf 39} (1985) 53.
J.~Ambjorn and S.~K.~Yang,
Phys.\ Lett.\ B {\bf 165} (1985) 140. T.~Matsui and A.~Nakamura,
Phys.\ Lett.\ B {\bf 194} (1987) 262.
H.~Okamoto, K.~Okano, L.~Schulke and S.~Tanaka,
Nucl.\ Phys.\ B {\bf 324} (1989) 684.
  K.~Okano, L.~Schulke and B.~Zheng,
  Phys.\ Lett.\ B {\bf 258} (1991) 421.
  L.~L.~Salcedo,
  Phys.\ Lett.\ B {\bf 305} (1993) 125.
  K.~Fujimura, K.~Okano, L.~Schulke, K.~Yamagishi and B.~Zheng,
  Nucl.\ Phys.\ B {\bf 424} (1994) 675
  [arXiv:hep-th/9311174].
  H.~Gausterer,
  J.\ Phys.\ A {\bf 27} (1994) 1325
  [arXiv:hep-lat/9312003].

\bibitem{Karsch:1985cb}
F.~Karsch and H.~W.~Wyld,
Phys.\ Rev.\ Lett.\  {\bf 55} (1985) 2242.
  J.~Flower, S.~W.~Otto and S.~Callahan,
  Phys.\ Rev.\ D {\bf 34} (1986) 598.
H.~Gausterer and J.~R.~Klauder,
Phys.\ Rev.\ D {\bf 33} (1986) 3678.
E.~M.~Ilgenfritz,
Phys.\ Lett.\ B {\bf 181} (1986) 327.
J.~Ambjorn and S.~K.~Yang,
Nucl.\ Phys.\ B {\bf 275} (1986) 18.
N.~Bilic, H.~Gausterer and S.~Sanielevici,
Phys.\ Lett.\ B {\bf 198} (1987) 235;
Phys.\ Rev.\ D {\bf 37} (1988) 3684.
  C.~W.~Bernard and V.~M.~Savage,
  Phys.\ Rev.\ D {\bf 64} (2001) 085010
  [arXiv:hep-lat/0106009].

\bibitem{Ambjorn:1986fz}
  J.~Ambjorn, M.~Flensburg and C.~Peterson,
  Nucl.\ Phys.\  B {\bf 275} (1986) 375;
  Phys.\ Lett.\  B {\bf 159} (1985) 335.

\bibitem{montvay} See e.g.\ I.~Montvay and G.~M\"unster, {\em Quantum Fields
on a Lattice} (Cambridge University Press, Cambridge, UK, 1997).

\bibitem{xue} S.~Xue, Phys.\ Lett.\ B {\bf 180} (1986) 275.

\bibitem{Grad}
I.~S.~Gradshteyn and I.~M.~Ryzhik, {\em Table of Integrals,
Series, and Products} (Academic Press, 5th ed., 1994).


\end{thebibliography}
\end{document}